\documentclass[fleqn,usenatbib]{mnras}

\usepackage{newtxtext,newtxmath}

\usepackage[T1]{fontenc}
\DeclareRobustCommand{\VAN}[3]{#2}
\let\VANthebibliography\thebibliography
\def\thebibliography{\DeclareRobustCommand{\VAN}[3]{##3}\VANthebibliography}

\usepackage{graphicx}	
\usepackage{amsmath}	

\title[NuSTAR Observations of OAO 1657-415]{Timing and Spectral Analysis of HMXB OAO 1657-415 with \emph{NuSTAR}}

\author[P. Sharma et al.]{
Prince Sharma,$^{1}$\thanks{E-mail: princerajsharma31@gmail.com}
Rahul Sharma,$^{1}$
Chetana Jain$^{2}$\thanks{E-mail: chetanajain11@gmail.com}
and Anjan Dutta$^{1}$
\\
$^{1}$Department of Physics and Astrophysics, University of Delhi, Delhi 110007, India\\
$^{2}$Hansraj College, University of Delhi, Delhi 110007, India\\
}

\date{Accepted XXX. Received YYY; in original form ZZZ}

\pubyear{2021}

\begin{document}
\label{firstpage}
\pagerange{\pageref{firstpage}--\pageref{lastpage}}
\maketitle

\begin{abstract}
This work presents a comprehensive timing and spectral analysis of high-mass X-ray binary pulsar, OAO 1657-415 by using the observation made with \emph{Nuclear Spectroscopic Telescope Array (NuSTAR)} on June 2019. During this observation, OAO 1657-415 exhibited X-ray variability by a factor of about 3. X-ray pulsations at 37.03322(14) s were observed up to 70 keV. OAO 1657-415 was undergoing a spin-down phase with $\dot{P} = 9(1) \times 10^{-8}$ s s$^{-1}$. This is an order of about 100 larger than the long-term spin-up rate. The pulse profile evolved marginally during the observation. We have discussed the long-term pulse period history of the source spanning  a time-base of 43 years, including the latest \emph{Fermi}/GBM data. The 3--70 keV source spectrum is described by a partially covered cutoff power-law, an Fe K$_{\alpha}$ line at 6.4 keV and a Compton shoulder at 6.3 keV. We report the presence of a cyclotron absorption feature around  40 keV, which is indicative of a surface magnetic field strength of $3.59 \pm 0.06 \ (1 + z)^{-1} \times 10^{12}$ and $3.29_{-0.22}^{+0.23} \ (1 + z)^{-1} \times 10^{12}$ G. This work shows the first robust presence of cyclotron absorption feature in the source.

\end{abstract}




\section{Introduction}
High-mass X-ray binaries (HMXBs) often host a neutron star (NS) that accretes matter lost from a massive companion star. The mode of mass transfer is either through stellar wind accretion or Roche-lobe overflow \citep{Reig2011,Falanga2015,Walter2015}. Based on the nature of the companion star, HMXBs can be classified into Be-X-ray binaries (BeXBs) or supergiant X-ray binaries (SgXBs). In BeXBs, NS accretes matter from the circumstellar accretion disc around the Be companion, whereas matter can be accreted either via stellar winds or Roche-lobe overflows in SgXBs \citep{Bildsten1997}.

OAO 1657-415 is an accretion powered HMXB pulsar discovered in 1978 with the \emph{Copernicus} satellite \citep{Polidan1978}. \citet{White1979} detected pulsations recurring at 38.22 s by using the \emph{HEAO} observations. \citet{Chakrabarty1993} reported its eclipsing nature by using the \emph{Compton Gamma-Ray Observatory} (\emph{CGRO}) data. OAO 1657-415 has an orbital period of 10.44 days with an eclipse lasting for 1.7 days and an eccentricity, $e \approx$ 0.104. The rate of orbital decay is estimated to be $\sim (-9.74 \pm 0.78) \times 10^{-8} $ \citep{Jenke2012}. The orbital parameters were consistent with a supergiant companion from spectral class B0--B6, having a mass between 14 and 18 M$_{\sun} $, and size in the range 25--32 R$_{\sun}$. A bright infrared counterpart was discovered in 2002 giving an estimated source distance of 6.4 $\pm$ 1.5 kpc \citep{Chakrabarty2002}. The infrared studies updated the spectral class of the companion to Ofpe/WNL \citep{Mason2009}.

The primeval timing analysis of OAO 1657-415 revealed a long-term spin-up with a rate $\dot{P} = -1.3 \times 10^{-9}$ s s$^{-1}$ \citep{White1979,Chakrabarty2002,Barnstedt2008}. Long-term monitoring showed large spin-up and spin-down episodes, similar to Cen X-3, on time-scales of few months up to a year \citep{Kamata1990,Chakrabarty1993,Bildsten1997}. \citet{Baykal1997} attributed this to be due to the formation of an episodic accretion disc. A marginal correlation of spin period and X-ray flux, during an extended spin-down phase, suggested the formation of disc in prograde direction \citep{Baykal2000}. The study of long-term spin period evolution with Burst and Transient Source Experiment \emph{CGRO}/BATSE and Gamma-Ray Burst Monitor \emph{Fermi}/GBM indicated two possible modes of accretion. The first mode favours the formation of a stable accretion disc (disk-wind accretion) resulting in a correlation between the spin-up and X-ray flux. In the other mode, NS spins down at a lower rate uncorrelated with flux during direct stellar wind accretion \citep{Jenke2012}.  

The wind-fed supergiant systems have longer orbital periods and longer spin periods because of weaker accretion torques while disc-fed supergiant systems show anti-correlation \citep{Corbet1986}. OAO 1657-415 occupies an intermediate position between these classes of supergiant systems \citep{Chakrabarty1993,Jenke2012}. 

The X-ray spectrum of OAO 1657-415 is highly absorbed due to its low galactic location and is generally described by using a power law model modified with either an exponential cutoff or a high energy cutoff component \citep{Kamata1990,Orlandini1999,Barnstedt2008}. Additionally, Fe emission lines at 6.4, 6.7, 6.97, 7.1 keV and Ni K$\alpha$ line at 7.4 keV have also been reported in the source spectrum \citep{Kamata1990,Orlandini1999,Audley2006,Pradhan2014,Pradhan2019,Jaiswal2021}. The study of \emph{ASCA} observation revealed the presence of a dust-scattered X-ray halo with decaying intensity through the eclipse and an estimated source distance of 7.1 $\pm$ 1.3 kpc.  \citet{Pradhan2014} studied the time-resolved spectrum of OAO 1657-415 by using the \emph{Suzaku} observation from 2011 and discussed the possibility of a clumpy stellar cloud as the reason for X-ray variability. \citet{Pradhan2019} have indicated the presence of Compton shoulder around 6.4 keV. 

Broad-band spectrum of OAO 1657-415 with \emph{BeppoSAX} indicated the presence of a Cyclotron Resonance Scattering Feature (CRSF) at $\sim$ 36 keV which was inconclusive due to limited statistics of the data \citep{Orlandini1999}. The \emph{INTEGRAL} spectrum did not show any CRSF signature \citep{Barnstedt2008}. The CRSF model parameters were poorly constrained even with the \emph{Suzaku} data \citep{Pradhan2014}.

In this paper, we report the results of timing and spectral analysis of HMXB OAO 1657-415 by using the \emph{NuSTAR} observation with an aim to study the pulse period evolution, presence of Compton scattered component and a possible CSRF feature in the source spectrum.  

\section{Observations}

\emph{Nuclear Spectroscopic Telescope Array (NuSTAR)} is NASA's first high-energy space mission devoted to X-ray imaging. It is equipped with advanced X-ray focusing telescopes that provide good sensitivity for imaging purposes above 10 keV \citep{Harrison2013}. It includes two identical detectors, known as focal plane module A (FPMA) and B (FPMB) fixed at the focus of each of the two co-aligned, grazing incidence telescopes with a field of view of 10 arcmin$^2$, an angular resolution of 18 arcsec, operating in 3--70 keV with an energy resolution (full-width at half-maximum) of 400 eV at 10 keV and 900 eV at 60 keV.

OAO 1657-415 was observed with \emph{NuSTAR} on 11 June 2019 for an exposure of 74.7 ks per FPM (Obs\_ID 30401019002). We carried out the standard data processing in \textsc{heasoft} v6.27 with \textsc{nustardas}\_01Apr20\_v1.9.2 and \emph{NuSTAR} \textsc{caldb} v20200912. We used the standard routine \textsc{nupipeline} to generate the screened and calibrated level 2 event files. The source events were extracted at the source position within a circular region of radius 90 arcsec for both FPMA and FPMB. A circular region of the same radius was used to extract the corresponding background events, from the source-free regions. We corrected the photon arrival time in the source event file to the solar system barycenter by using the source coordinates R.A. $= 255.20368^\circ$ and Dec. $= -41.65596^\circ$ \citep{Cat2003}. We generated the spectrum, light curves, and corresponding response files by using the task \textsc{nuproducts}. 
\begin{figure}
\centering
	\includegraphics[width=\columnwidth]{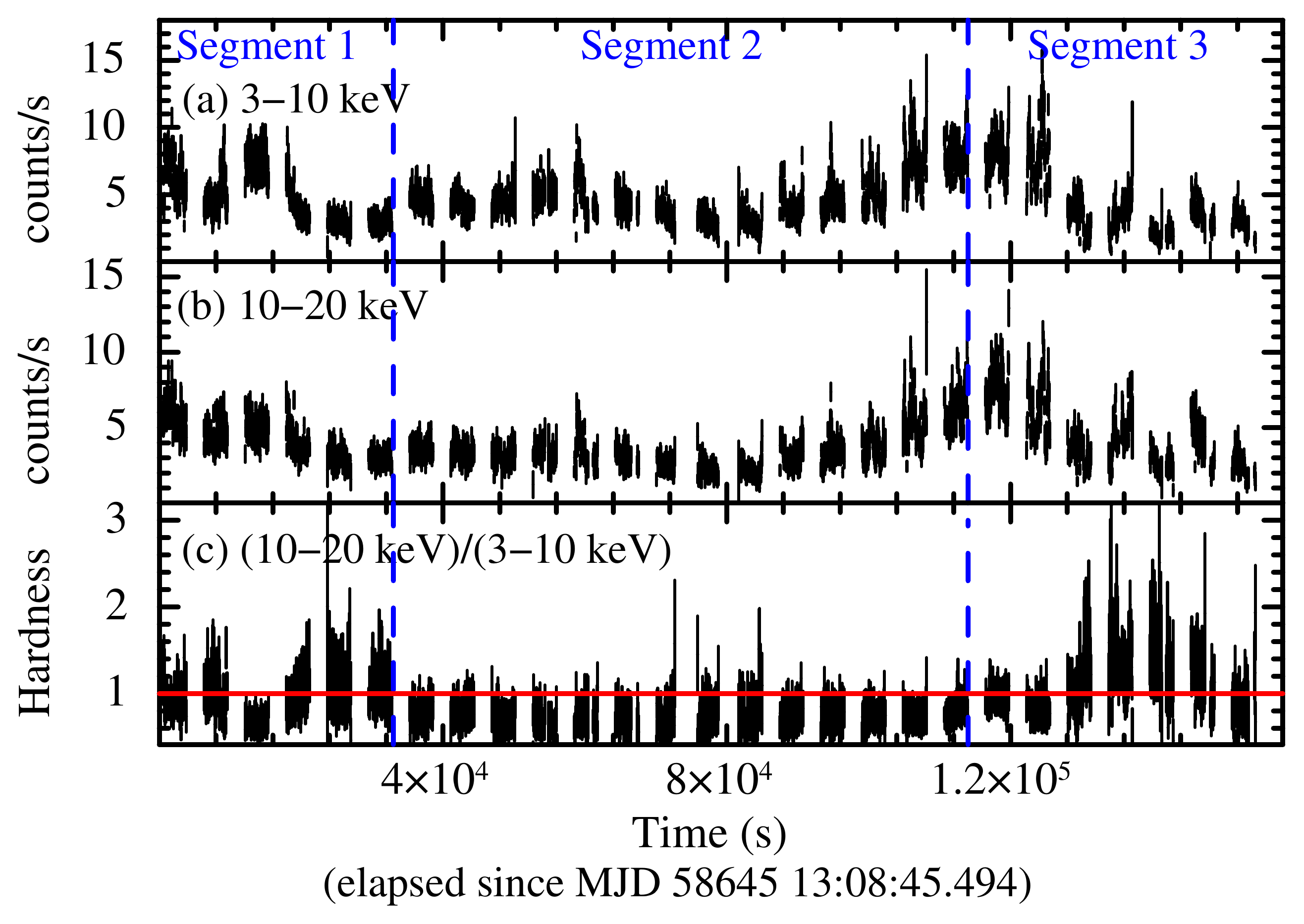}
    \caption{Background-subtracted light curve of OAO 1657-415 binned at 32 s. The top and middle panels show light curves in 3--10 keV and 10--20 keV, respectively. The bottom panel gives the hardness ratio defined as the ratio between 10--20 and 3--10 keV count rate. The entire observation is divided into three segments marked by dashed vertical lines.} 
    \label{fig:longlc}
\end{figure}

Figure \ref{fig:longlc} shows the background-subtracted light curve of OAO 1657-415 for the entire observation. The top and middle panels correspond to energy range 3--10 keV and 10--20 keV, respectively. The hardness ratio (HR) (ratio of count rate in 10--20 keV and 3--10 keV) is given in the bottom panel. The X-ray flux varied significantly during this observation. Therefore, for the timing and spectral analysis, we have divided the entire observation into three segments, as shown by vertical dashed lines in Figure \ref{fig:longlc}. The first segment corresponds to MJD 58645.55 to 58645.92, the second goes from MJD 58645.95 to 58646.85 and the third one is during MJD 58646.85 to 58647.33.

\begin{figure}
\centering
	\includegraphics[width=\columnwidth]{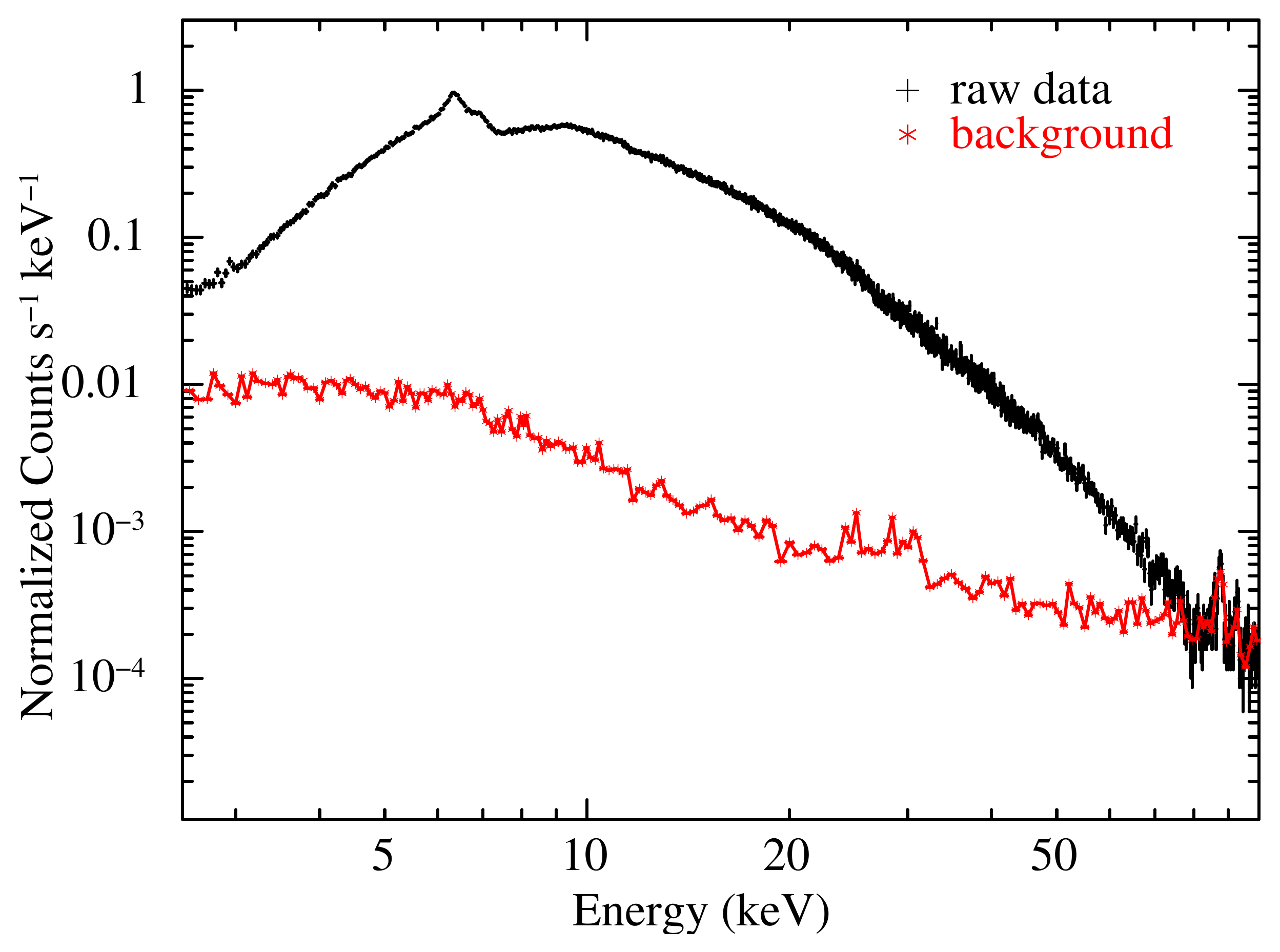}
    \caption{Raw source and background spectra of OAO 1657-415 observed with \emph{NuSTAR}. Background count rate dominates above 70 keV.} 
    \label{fig:back}
\end{figure}

The source was detected up to 70 keV above background during the entire observation (Figure \ref{fig:back}). We limited our spectral analysis between the energy range 3 and 70 keV. We added the two source spectra, background spectra and response files from FPMA and FPMB by using \textsc{addspec} tool to improve the statistics of the data, for all three segments. We re-binned all three spectra to contain at least 25 counts per energy bin.

\section{Results}
\subsection{Timing analysis}
\label{sec:timing} 

\begin{figure}
\centering
	\includegraphics[width=\columnwidth]{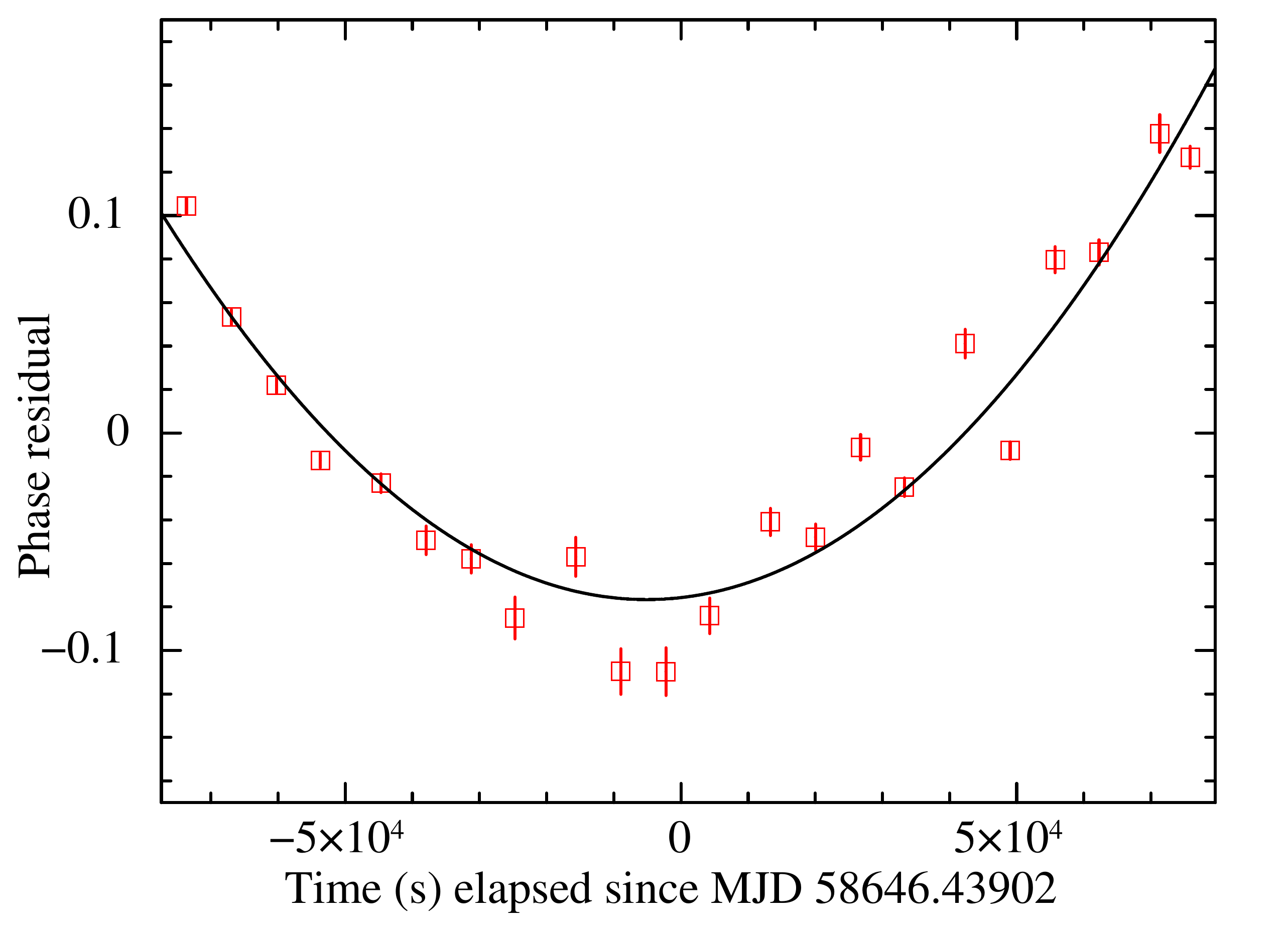}
    \caption{Measured timing residual after fitting the pulse phase with period P = 37.031301(1) s in 3--70 keV. The solid curve shows the best-fitting function up to a quadratic term.} 
    \label{fig:phase}
\end{figure}

The light curve of OAO 1657-415 covering the entire observation showed the variation in X-ray flux up to 70 keV energy range. As observed in Figure \ref{fig:longlc}, the count rate and HR was relatively higher for the first 32 ks (Segment 1) as compared to that in subsequent 78 ks (Segment 2). During the last 40 ks (Segment 3), the count rate and HR roughly increased to initial values. 

We searched the entire light curve in 3--70 keV band for periodicity by using the \textsc{powspec} and \textsc{efsearch} tools of FTOOLS. We obtained a pulse period of 37.031301(1) s. For the phase-coherent timing, we divided the light curve into 22 intervals of 5800 s each (equal to roughly 1 \emph{NuSTAR} orbit) and generated the pulse profiles corresponding to each interval. The pulse profile from each interval was cross-correlated with a master pulse profile generated from the entire light curve. The phase delay from each interval with respect to the master profile is shown in Figure \ref{fig:phase}. Timing solution was obtained by fitting Equation \ref{eq:phi} to the phase residual trend \citep{Baykal2000}. We obtained a $\chi^2$ of 290 for 19 degrees of freedom (dof). Since the data points are scattered about the best-fitting curve, we rescaled the errors on the best-fitting parameters by multiplying them with the square root of the reduced $\chi^2$. This gave a pulse period of 37.03322(14) s and a period derivative of $\dot{P} = 9(1) \times 10^{-8}$ s s$^{-1}$.
\begin{equation}
 \phi (t) = \phi_0 + \delta \nu ( t - t_0 ) + \dot{\nu}/2 ( t - t_0 )^2
\label{eq:phi}
\end{equation}

\begin{table}
	\caption{Measured value of pulse period of OAO 1657-415. The epoch corresponds to the mid point of the observation interval.}
	\label{tab:period}
	\begin{tabular*}{\columnwidth}{ccccc}
	\hline
Seg & Epoch & Exposure$^{a}$ & Count Rate$^{b}$ &Pulse Period \\ 
		 & (MJD) & (ks) & (count s$^{-1}$) & (s) \\ 
		\hline
		1 & 58645.74 & 36.04 & $6.985 \pm 0.014$ & 37.024578(5) \\[0.5ex] 
		2 & 58646.40 & 74.60 & $6.550 \pm 0.009$ & 37.031682(7) \\[0.5ex] 
		3 & 58647.09 & 36.76 & $7.673 \pm 0.014$ & 37.036472(7)  \\[0.5ex] 
	\hline
	\multicolumn{5}{l}{\textit{Notes.} $^{a}$Combined exposure for the \emph{NuSTAR} FPMA and FPMB spectra.}\\
    \multicolumn{5}{l}{$^{b}$Average count rate in 3--70 keV energy range.}\\
	\end{tabular*}
\end{table}
We extracted the individual background-corrected light curves corresponding to each of the three segments for our timing analysis. We generated the power density spectra (PDS) for all three segments. We used the root-mean-square (rms) normalization and re-binned the PDS geometrically by a factor of 1.03 to improve the signal to noise ratio of frequency bins. We detected strong peaks in each PDS at around 0.026 Hz. Table \ref{tab:period} lists the exposure, pulse period, and corresponding epoch for each segment. The pulse period increased gradually from first to last segment. The evolution of spin period is consistent with the spinning-down of the pulsar.

\begin{figure*}
   \includegraphics[width=\columnwidth]{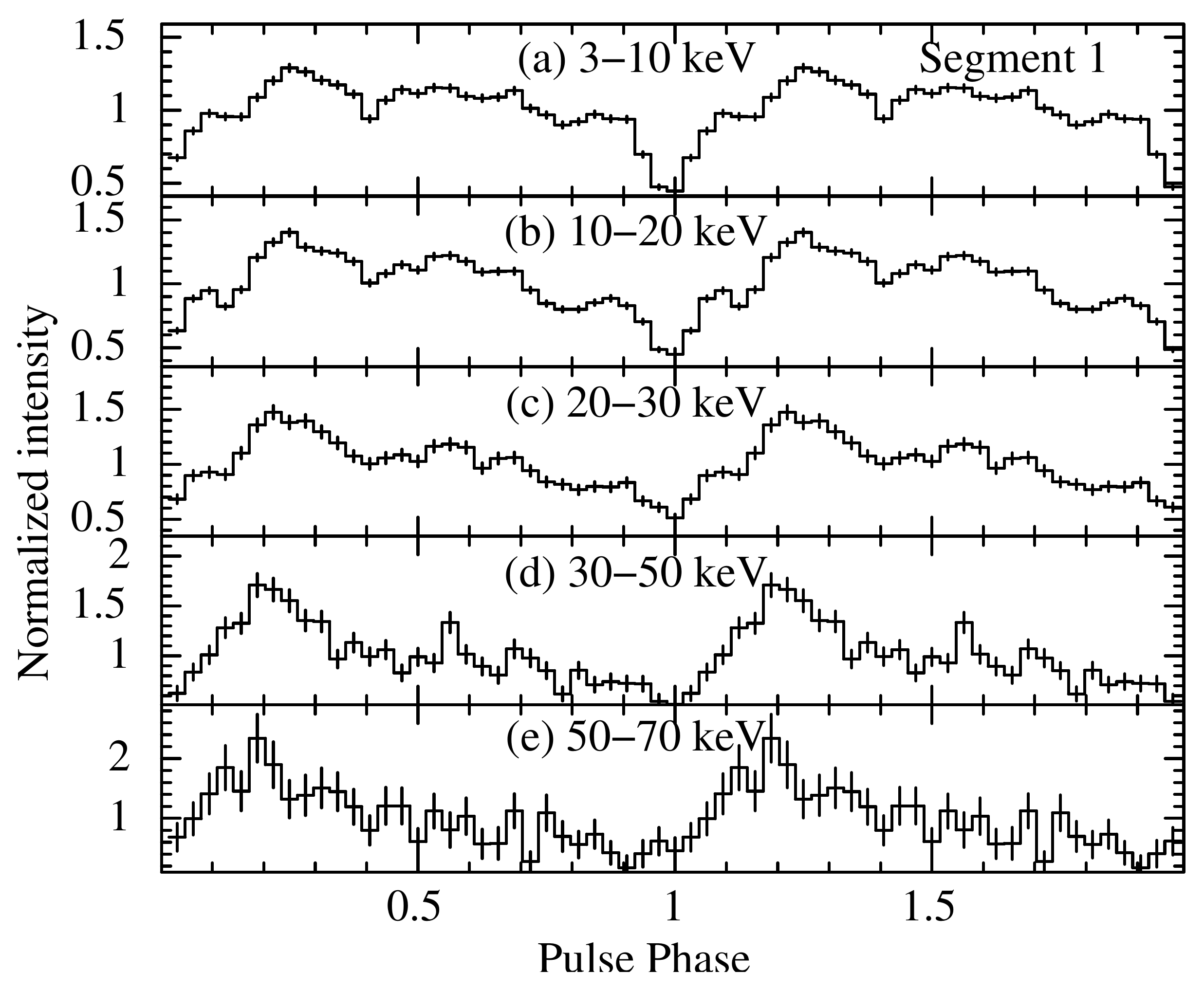}
   \includegraphics[width=\columnwidth]{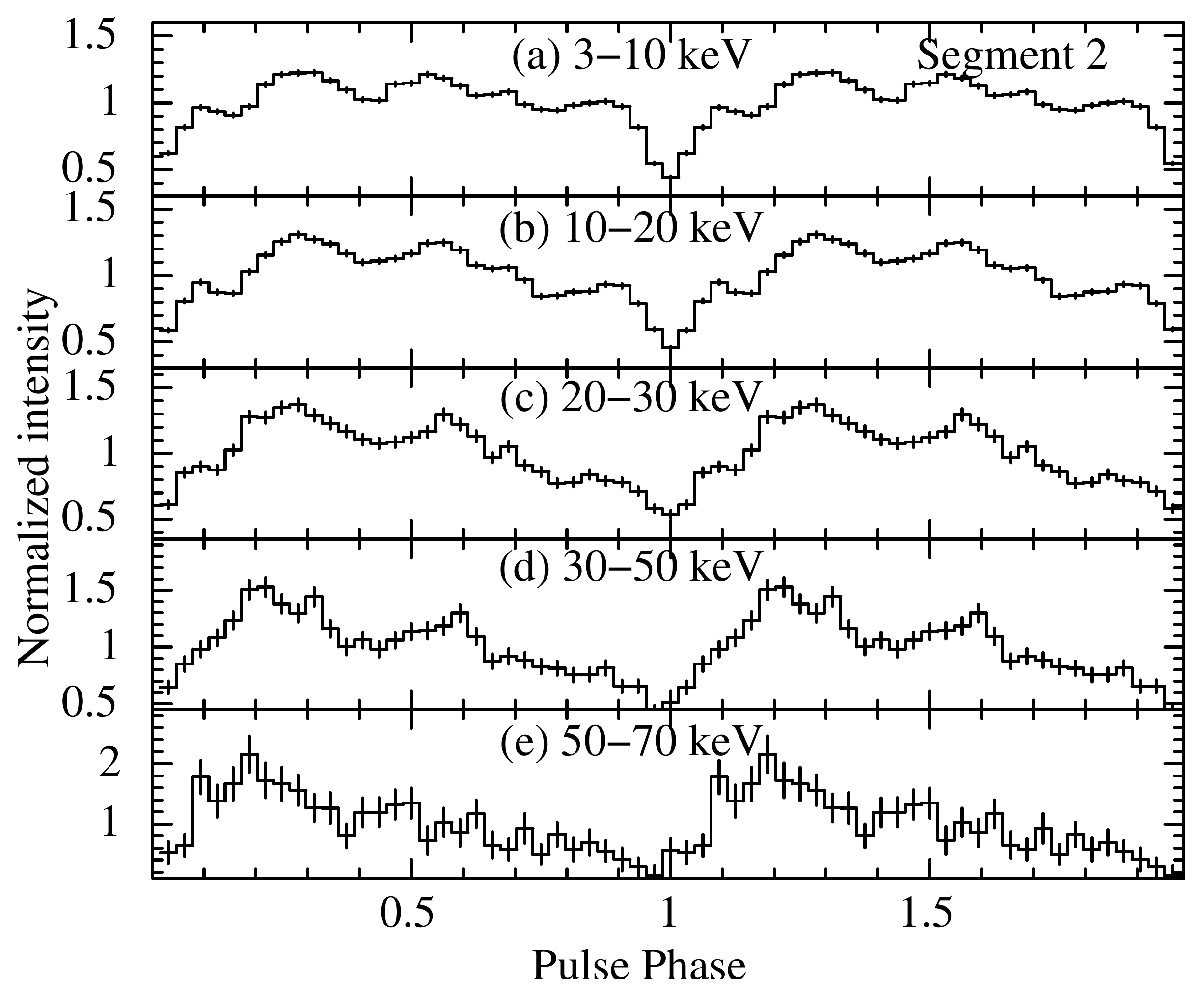}
   \includegraphics[width=\columnwidth]{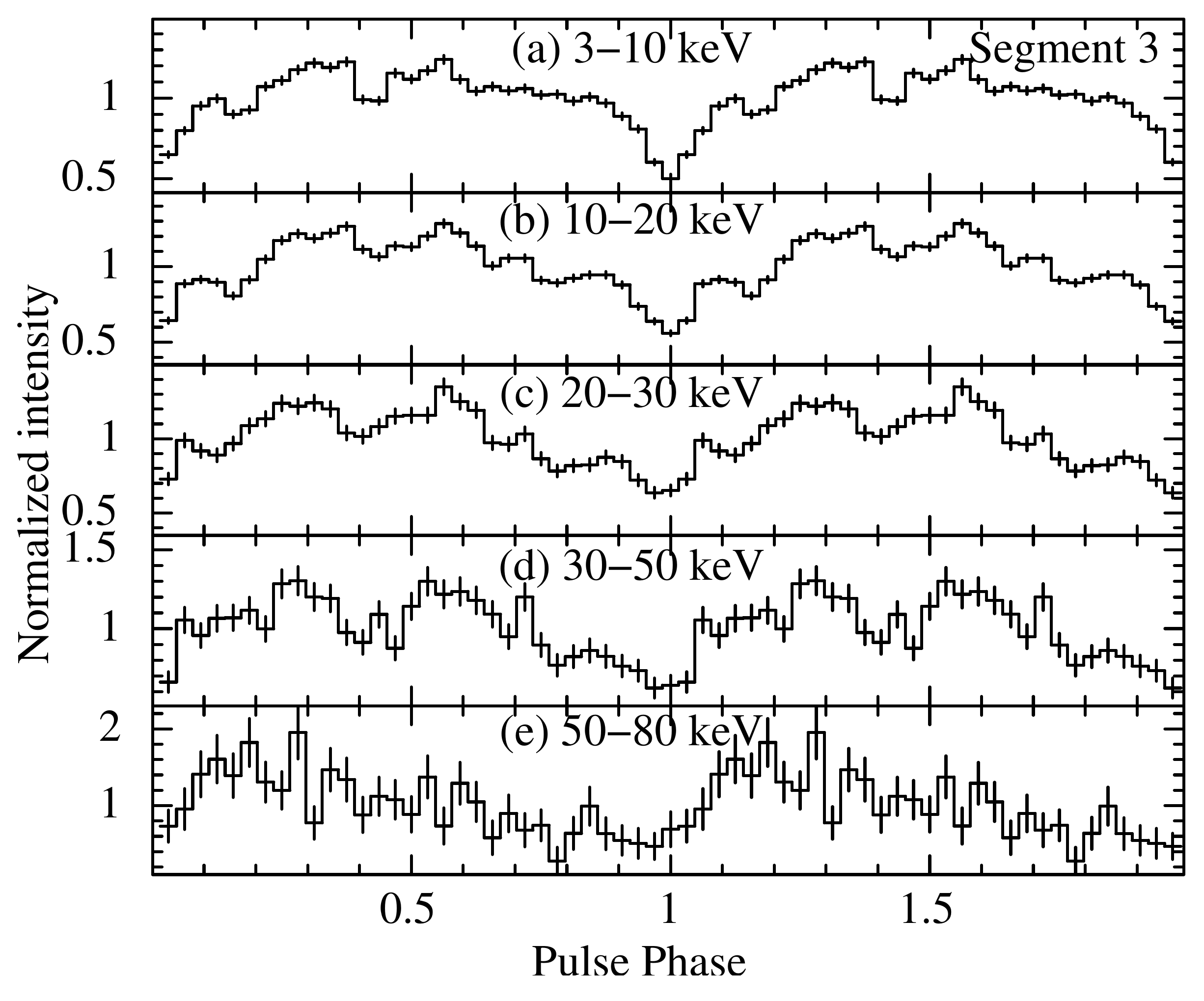}
\caption{Energy resolved pulse profiles obtained from epoch-folding of \emph{NuSTAR} light curves for segment 1,2, and 3. The profiles are plotted for two cycles for clarity.}
 \label{fig:profile}
\end{figure*}

We folded the segmented light curves by using the respective best periods to generate the energy-resolved pulse profiles in 3--10, 10--20, 20--30, 30--50, and 50--70 keV energy ranges as shown in Figure \ref{fig:profile}. The pulse profiles were observed to evolve with time and energy. For segment 1, it was observed that the dip near the main peak broadened with energy. The profile shows a secondary peak around 0.5 pulse phase for 30--50 keV energy band. For 50--70 keV energy band the main peak becomes sharper with a flat tail. The shape of the pulse is moderately different during the second segment with a broader dip near the main peak and shows the presence of a secondary peak above 20 keV. With increasing energy, the profile becomes double-peaked with additional tiny feature close to the first peak. During the last segment, dip in the pulse shape becomes more pronounced which increases with energy. The overall shape of the profile is different from that during the earlier segments above 30 keV energy with a stronger secondary peak for 30--50 keV.  

\begin{figure}
\centering
	\includegraphics[width=\columnwidth]{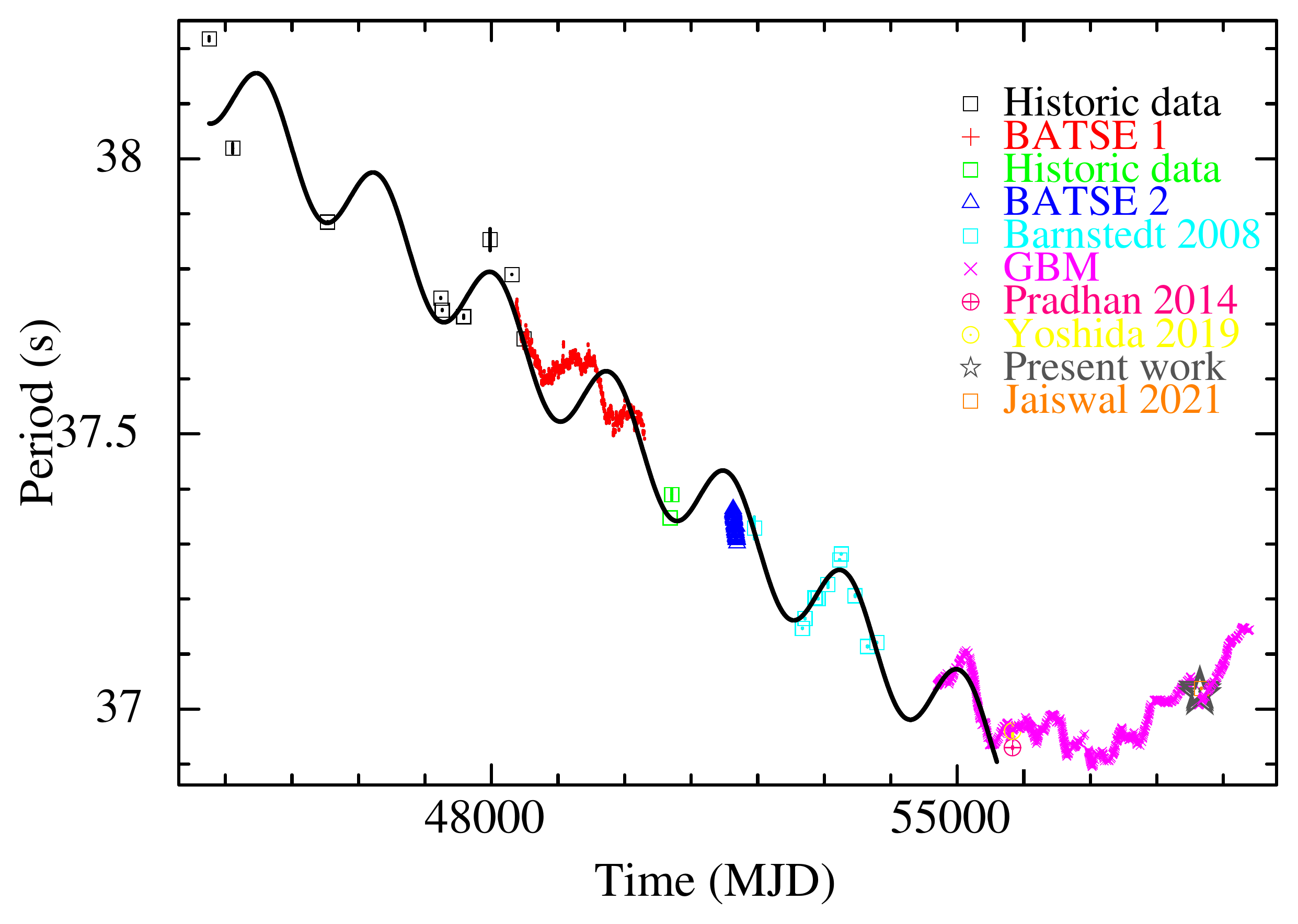}
	\includegraphics[width=\columnwidth]{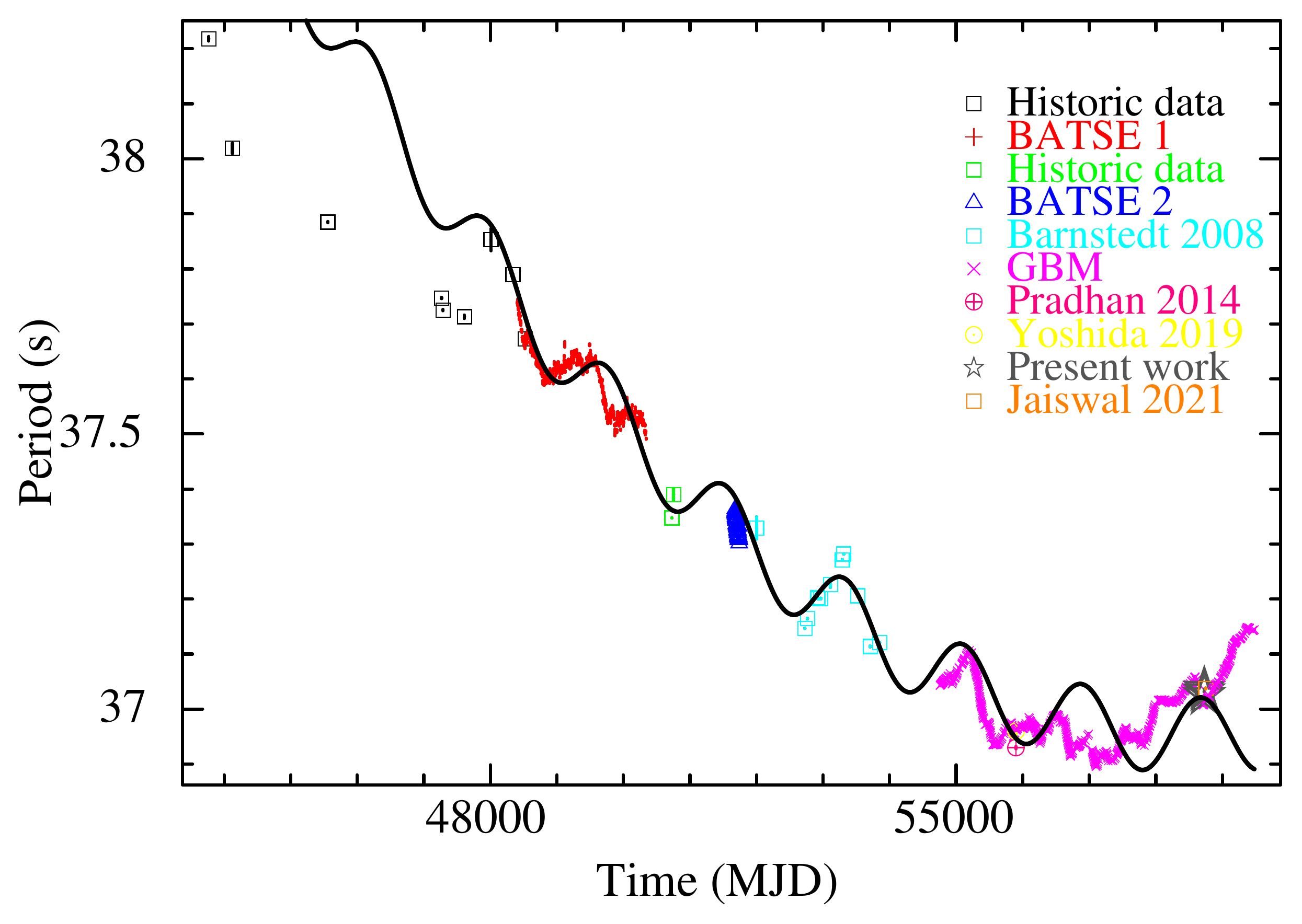}
	\includegraphics[width=\columnwidth]{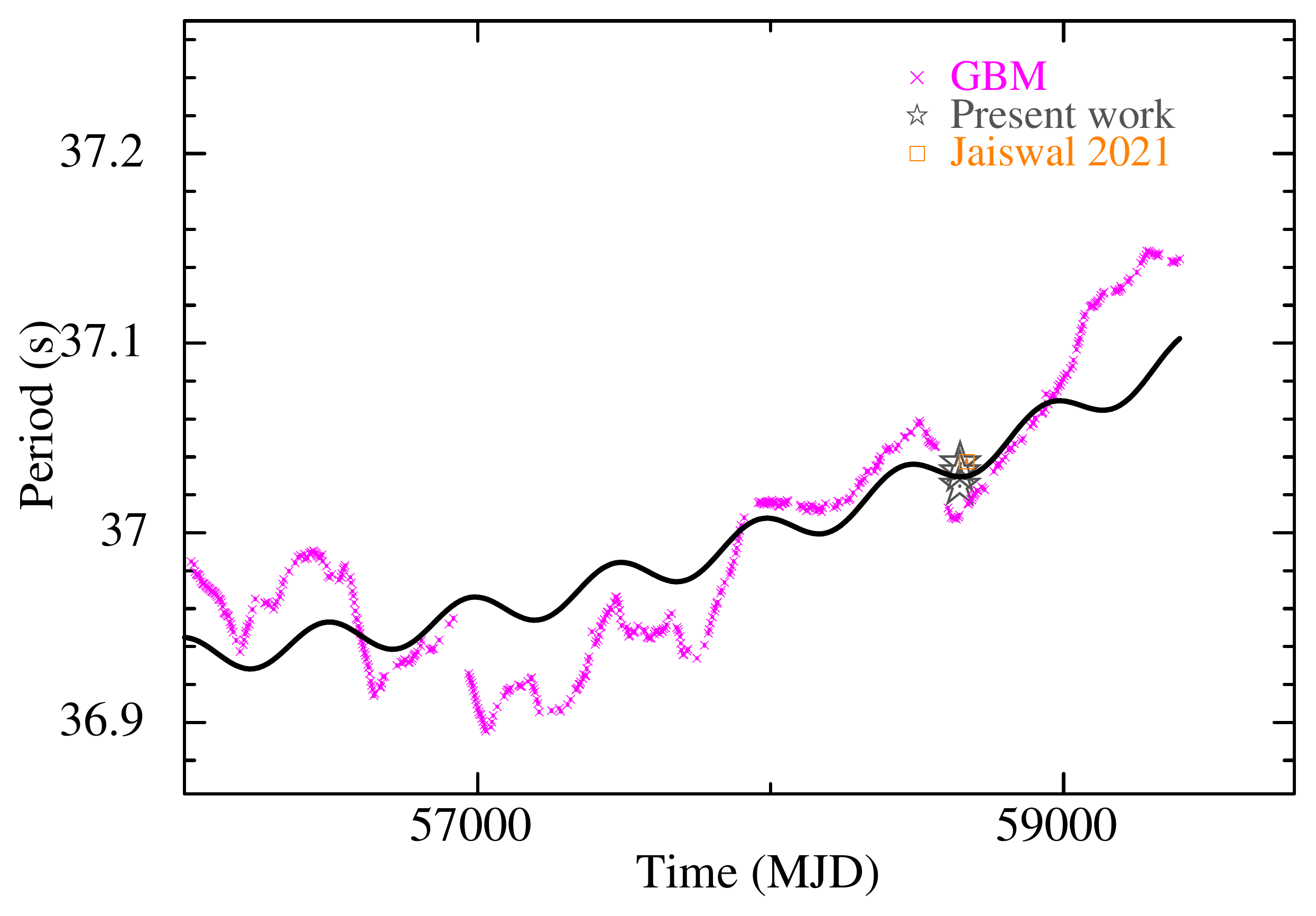}

    \caption{Long-term spin period history of OAO 1657-415 covering 43 years between 1978 and 2021. The different markers represent data from different sources. Historic data marked with black $\square$ include results from \citet{White1979, Parmar1980, Nagase1984, Kamata1990, Mereghetti1991} and \citet{Sunyaev1991}. The data represented by red + and blue $\bigtriangleup$ indicates \emph{CGRO}/BATSE data available from the public archive. Historic data marked with green $\square$ represent results from \citet{Baykal2000} and \citet{Audley2006}. The \emph{Fermi}/GBM archival data are marked with pink $\times$ marker. Grey $\star$ marks the result from the present work. Solid curve in the first plot corresponds to the best-fitting linear-sinusoidal model for MJD 43755 and 56000. The solid curves in the second and third plot correspond to the best-fitting model comprising of quadratic and sinusoidal functions}.   
    \label{fig:phistory}
\end{figure}

\subsection{Long-term spin evolution and torque reversal}

The long-term spin period evolution of OAO 1657-415, covering three decades between 1978 and 2006, by using the data from several missions was presented by \citet{Barnstedt2008}. Long-term history showed variations in the spin period on a time-scale of about 4.8 yr superimposed on an overall secular spin-up. In order to study the period evolution, we extended the time-base by 15 years by including the numbers from reported works, \emph{Fermi}/GBM \citep{Meegan2009} and the present work. Figure \ref{fig:phistory} shows the spin period history of OAO 1657-415 starting from September 1978 up to July 2021. The spin-up trend lasted for around 32 years since the discovery of the source with recurring episodes of spin-up/down on shorter time-scales.

The short-term fluctuations in the spin period are known to be a characteristic property of wind-fed pulsars \citep{Deeter1989,Bildsten1997}. The deviation from an initial spin-up trend happened around MJD 47600 lasting for about 1 yr before entering the spin-up phase again. The next spin-up phase lasted for about 2.3 yr (MJD 48900). Similar reversals in the spin period trend are evident at around MJD 49500, 53200, and 55100. The long-term spinning up of the pulsar continued up to around MJD 56000. Beyond around MJD 57300, small fluctuations overlying on an overall spin-down phase are clearly seen. The source has been in this phase for around 6 yr now.

In order to quantify the presence of short-term periodicity, we modeled the long-term spin period history of OAO 1657-415 with linear and quadratic functions, each modulated with a sinusoidal function. The solid curve in the first plot of Figure \ref{fig:phistory} corresponds to the best-fitting linear and sinusoidal function for spin period history up to around MJD 56000. The best-fitting gave a sinusoidal periodicity of 1744 d with an amplitude of 80 ms and a long-term spin-up rate of $\dot{P} = -1.2 \times 10^{-9}$ s s$^{-1}$. The parameter values are consistent with the results of \citet{Barnstedt2008}. A model comprising of quadratic and a sinusoidal function was fit to the entire spin history of the source. This gave a sinusoidal period of 1770 d, amplitude of 70 ms and spin period derivative of $\dot{P} = -3 \times 10^{-9}$ s s$^{-1}$. As seen in the middle plot of Figure \ref{fig:phistory}, the fit is relatively poor before MJD 48000 and after MJD 56000. It is clear that the spin period evolution in OAO 1657-415 changes remarkably around MJD 56000. A separate fit of quadratic plus sinusoidal model to data points after MJD 56000 yielded a period of about 492 d with period variation amplitude of about 10 ms (bottom plot of Figure \ref{fig:phistory}). We conclude that OAO 1657-415 continued to show short-term variations with a period of 1744 d (4.8 yr) up to around MJD 56000. We also surmise that OAO 1657-415 has undergone a permanent switch from spinning up to spinning down around MJD 57300 (May 2015). A similar torque reversal has been reported in X-ray pulsars Vela X-1 \citep{Hayakawa1982} and GX 1+4 \citep{Gonzalez2012}.

\subsection{Spectral Analysis}

We have utilized the spectral analysis package \textsc{xspec} v12.11.0k \citep{Arnaud1996}, distributed with the \textsc{heasoft} package, to model the \emph{NuSTAR} spectra of OAO 1657-415 and have used the $\chi^2$ statistics as the test statistics. We have adopted  solar abundances from \citet{Wilms2000} and the photoelectric cross-sections from \citet{Verner1996}. The lower and upper limits for parameter uncertainties are reported at 90 per cent confidence level.

X-ray continua of accreting HMXB pulsars are generally described by a power law modified with a high-energy cutoff in the range 10--30 keV along with a soft blackbody component \citep{White1983,Coburn2002,Becker2005}. We first fit the spectra from three segments simultaneously, with all parameters free to vary, by using a cutoff power law (\texttt{cutoffpl}) component. We included \texttt{tbabs} to account for the photoelectric absorption due to the interstellar medium along the direction of the source. Values for the column density in the three segments were found to be similar (Segment 1: $5.33_{-4.84}^{+6.00} \times 10^{22}$ , Segment 2: $< 8.26 \times 10^{22}$, Segment 3: $12.35_{-2.76}^{+4.73} \times 10^{22}$ cm$^{-2}$). The ${N_{\rm H}}$ values from previous works show a large range from $1.29 \pm 0.04 \times 10^{22}$ to $75.74 \pm 5.91 \times 10^{22}$ cm$^{-2}$ \citep{Orlandini1999,Baykal2000,Audley2006,Jaiswal2014,Pradhan2019}. So, for the subsequent analysis, we have fixed ${N_{\rm H}} = 5 \times 10^{21}$ cm$^{-2}$, the lowest value obtained for the first segment. The model failed to provide an adequate fit to the spectra with large positive residual at the lower energy end. We also found systematic residuals around 6.4 keV, possibly due to the Fe K$_{\alpha}$ emission line common in the spectra of HMXBs. The inclusion of a blackbody component to account for the soft excess and a Gaussian component to model the emission like feature between 6--7 keV improved the fit but excess residual below 4 keV was still present. The best-fitting Gaussian line occurred at 6.3 keV. 

\begin{figure}
   \includegraphics[width=\columnwidth]{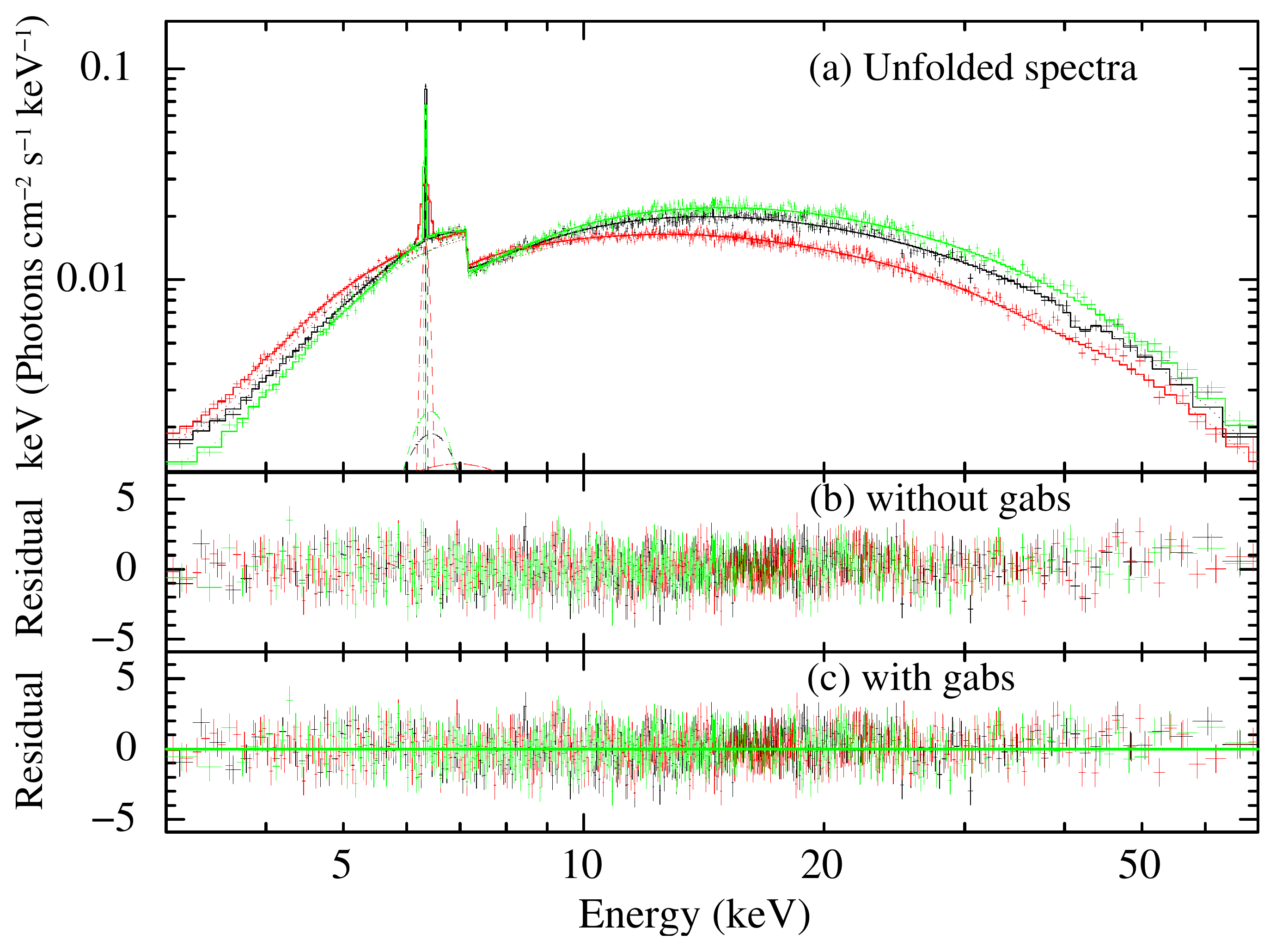}
\caption{(a) The unfolded \emph{NuSTAR} spectra and model component for segment 1 (black), segment 2 (red), and segment 3 (green) modelled with \texttt{tbabs*gabs*(cutoffpl*pcfabs + gaus + gaus)}. (b) Residual before addition of \texttt{gabs}. (c) Residual after adding \texttt{gabs} component.}
 \label{fig:spec1}
\end{figure}

We then removed the blackbody component and included a partially covering absorption component (\texttt{pcfabs}) to model the spectra. The model fit the spectra well and removed the low energy residual returning a $\chi^2$ of 3563 for 3039 dof. The best-fitting Gaussian emission line occurred at $6.34 \pm 0.01$, $6.34 \pm 0.01$, and $6.33 \pm 0.01$ keV for three spectra, respectively. The residual showed a narrow feature around $\sim$ 6.5 keV. We added another Gaussian component and constrained the line energy between 6.4 and 7 keV. This model \texttt{tbabs*(cutoffpl*pcfabs + gaus + gaus)} provided an adequate fit with a $\chi^2$/dof = 3395/3033. We obtained upper limits of 6.43, 6.56, and 6.42 keV for three spectra, respectively, for the line energy of the second Gaussian component. We identify this feature as Fe K$_{\alpha}$ emission line whereas the 6.34 keV feature resembles the Compton shoulder. Figure \ref{fig:spec1} shows the best-fitting spectra along with the individual model components for all three segments. Finally, we also noticed a narrow absorption like dip in the residual around 40 keV in the spectra of segments 1 and 2 (Figure \ref{fig:spec1}(b)). 

\begin{figure}
   \includegraphics[width=\columnwidth]{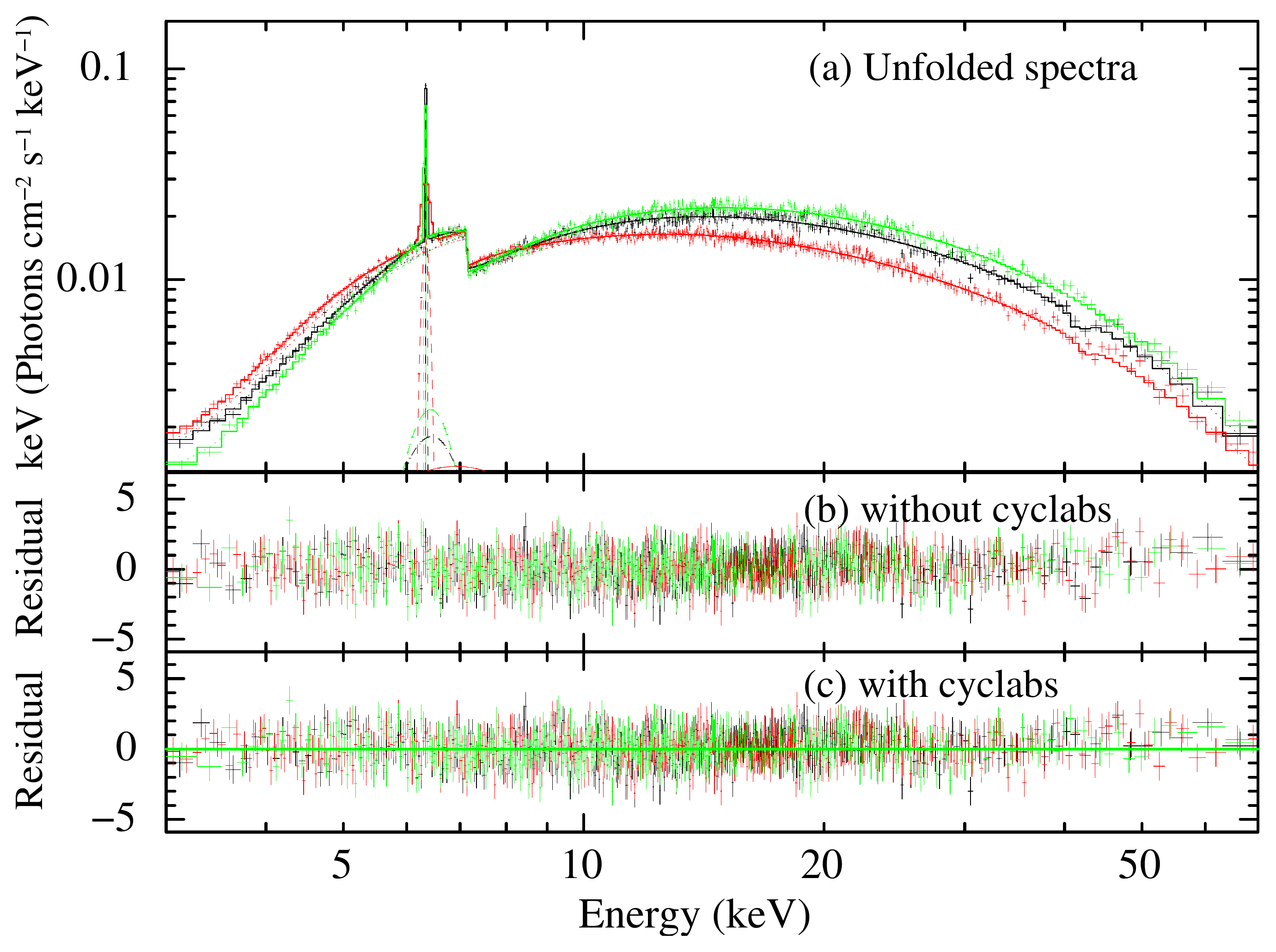}
\caption{(a) The unfolded \emph{NuSTAR} spectra and model component for segment 1 (black), segment 2 (red), and segment 3 (green) modelled with \texttt{tbabs*cyclabs*(cutoffpl*pcfabs + gaus + gaus)}. (b) Residual before addition of \texttt{cyclabs}. (c) Residual after adding \texttt{cyclabs} component.}
 \label{fig:spec2}
\end{figure}

Presence of CRSF around $\sim$ 36 keV in the broad-band X-ray spectra of OAO 1657-415 obtained with \emph{BeppoSAX} and \emph{Suzaku} have been suggested by previous studies \citep{Orlandini1999,Pradhan2014}. However, the detection remained inconclusive due to the statistical limitation of data. We tried to model this feature by using a Gaussian absorption line component (\texttt{gabs}) and cyclotron absorption line component (\texttt{cyclabs}). The inclusion of either of the absorption components improved the fit by reducing $\chi^2$ by 23 and 18 for 6 additional parameters for \texttt{gabs} and \texttt{cyclabs}, respectively. Both the models provided a suitable fit with similar parameter values. We report results from both models. Figures \ref{fig:spec1} and \ref{fig:spec2} display the best-fitting spectra from three segments and residuals before and after including the absorption components. Table \ref{tab:model} reports the best-fitting spectral parameters for two models.

\begin{table*}
\caption{Best-fitting spectral parameters for \emph{NuSATR} spectra of OAO 1657-415 from three segments. The errors are given at 90 per cent confident level.}
\label{tab:model}

\begin{tabular*}{2\columnwidth}{|c c | c c c |c c c|} 
\hline
\textbf{Component} & \textbf{Parameters} & \multicolumn{3}{c}{\textbf{Model: tbabs*gabs*(pcfabs*cutoffpl + gaus + gaus)}} & \multicolumn{3}{c}{\textbf{Model: tbabs*cyclabs*(pcfabs*cutoffpl + gaus + gaus)}}\\
\hline
& & \textbf{Segment 1} & \textbf{Segment 2} & \textbf{Segment 3} &\textbf{Segment 1} & \textbf{Segment 2} & \textbf{Segment 3} \\
\hline

 PCFABS & ${N_{\rm H2}}^a$ & $ 58_{-3}^{+2}2 $ & $ 52_{-2}^{+1} $ & $ 60_{-3}^{+1} $ & $ 59 \pm 3 $ & $ 53_{-2}^{+1} $ & $ 60_{-3}^{+2} $ \\[0.5ex]
 & $Cvr.f$ & $ 0.92 \pm 0.01 $ & $ 0.93 \pm 0.01 $ & $ 0.94 \pm 0.01 $ & $ 0.92 \pm 0.01 $ & $ 0.93 \pm 0.01 $ & $ 0.93 \pm 0.01 $\\[0.7ex]

 CUTOFFPL & $\Gamma$  & $ 0.49_{-0.04}^{+0.08} $ & $ 0.80_{-0.03}^{+0.02} $ & $ 0.36_{-0.08}^{+0.04} $ & $ 0.50 \pm 0.07 $ & $ 0.80 \pm 0.04 $ & $ 0.35_{-0.04}^{+0.07} $\\[0.5ex]
 & $E_{\rm cut}$ (keV) & $ 15.61_{-0.41}^{+0.90} $ & $ 18.82_{-0.45}^{+0.38} $ & $ 14.82_{-0.35}^{+0.66} $ & $ 15.74_{-0.72}^{+0.41} $ & $ 18.54_{-0.61}^{+0.54} $ & $ 14.78_{-0.54}^{+0.78} $\\[0.5ex]
 & Norm ($10^{-2}$) & $ 1.46_{-0.24}^{+0.28} $ & $ 2.32_{-0.24}^{+0.27} $ & $ 1.20_{-0.19}^{+0.10} $ & $ 1.49_{-0.23}^{+0.27} $ & $ 2.35_{-0.22}^{+0.25} $ & $ 1.19_{-0.08}^{+0.21} $\\[0.5ex]
 & ${f_{\rm bol}}^b$ & $10.37 \pm 0.06 $ & $8.50  \pm 0.04 $ & $11.54 \pm 0.06 $ & $ 10.39 \pm 0.06 $ & $ 8.43 \pm 0.04 $ & $ 11.54 \pm 0.06 $  \\[0.7ex]
 
 GAUS & $E_{\rm line}$ (keV) & $ 6.36_{-0.02}^{+0.01} $ & $ 6.34 \pm 0.01 $ & $ 6.32_{-0.01}^{+0.11} $ & $ 6.36_{-0.05}^{+0.01} $ & $ 6.34 \pm 0.01 $ & $ 6.32_{-0.01}^{+0.03} $ \\[0.5ex]
 & $\sigma$ (keV) & $ < 0.06 $ & $ 0.07_{-0.03}^{+0.02} $ & $ < 0.06 $ & $ < 0.06 $ & $ 0.07_{-0.03}^{+0.02} $ & $ < 0.05 $\\ [0.5ex]
 & Norm($10^{-4}$) & $ 4.65_{-0.42}^{+0.44} $ & $ 4.44 \pm 0.26 $ & $ 4.47_{-0.28}^{+0.46} $ & $ 4.68_{-0.44}^{+0.41} $ & $ 4.40_{-0.25}^{+0.26} $ & $ 4.44_{-0.47}^{+0.52} $ \\[0.7ex]

 GAUS & $E_{\rm line}$ (keV)  & $ < 6.43 $ & $ < 6.58 $ & $ < 6.42 $ & $ < 6.43 $ & $ < 6.58 $ & $ < 6.42 $\\[0.5ex]
 & $\sigma$ (keV) & $ 0.56_{-0.10}^{+0.12} $ & $ 1.99_{-0.21}^{+0.16} $ & $ 0.43_{-0.06}^{+0.08} $ & $ 0.57_{-0.10}^{+0.12} $ & $ 1.95_{-0.21}^{+0.18} $ & $ 0.42_{-0.06}^{+0.08} $\\ [0.5ex]
 & Norm($10^{-4}$) & $ 4.10_{-0.74}^{+0.71} $ & $ 10.03_{-2.26}^{+1.62} $ & $ 3.98_{-0.64}^{+0.62} $ & $ 4.02_{-0.67}^{+0.68} $ & $ 9.57_{-2.19}^{+1.80} $ & $ 4.04_{-0.65}^{+0.60} $\\[0.5ex]
 & Eqw(eV) & $ 144_{-32}^{+49} $ & $ 429_{-274}^{+193} $ & $ 160_{-51}^{+41} $ & $ 141_{-31}^{+48} $ & $ 165_{-19}^{+419} $ & $ 124_{-9}^{+176} $\\[0.5ex]
 & ${f_{\rm bol}}^c$ & $4.20 \pm 0.43 $ & $10.28 \pm 0.54 $ & $ 4.08 \pm 0.38 $ & $ 4.12 \pm 0.43 $ & $ 9.76 \pm 0.53 $ & $ 4.13 \pm 0.38 $\\[0.7ex]

GABS & $E_{\rm abs}$ (keV)  & $ 41.68_{-0.72}^{+0.69} $ & $ 38.13_{-2.49}^{+2.74} $ & - & & & \\[0.5ex]
 & $\sigma$ (keV) & $ 0.86_{-0.46}^{+1.02} $ & $ 5.28_{-1.60}^{+2.34} $ & - & & & \\ [0.5ex]
 & Depth & $ 0.40_{-0.21}^{+0.27} $ & $ 0.85_{-0.34}^{+0.49} $ & - & & &\\[0.7ex]
 
 CYCLABS & $E_{\rm abs}$ (keV)  &  &  &  & $ 41.61_{-0.69}^{+0.72} $ & $ 42.42_{-1.10}^{+0.96} $ & - \\[0.5ex]
 & Width (keV) &  &  &  & $ 0.86_{-0.62}^{+1.64} $ & $ < 3.02 $ & - \\ [0.5ex]
 & Depth &  &  &  & $ 0.20_{-0.12}^{+0.24} $ & $ 0.12_{-0.08}^{+0.07} $ & - \\[0.7ex]

&${f_{\rm Total}}^b$ & $10.46 \pm 0.06 $ & $8.65 \pm 0.03 $ & $11.63 \pm 0.06 $ & $ 10.48 \pm 0.06 $ & $ 8.57 \pm 0.03 $ & $ 11.62 \pm 0.06 $ \\[0.5ex]
 \hline
 & $\chi^2$/dof & 1108/945 & 1174/1079 & 1090/1003 & 1108/945 & 1179/1079 & 1090/1003 \\
\hline
\multicolumn{5}{l}{\textit{Notes.} $^a$ In units of $10^{22}$ cm$^{-2}$}\\
\multicolumn{5}{l}{$^b$ is 0.1--100 keV unabsorbed bolometric flux in units of $10^{-10}$ erg cm$^{-2}$ s$^{-1}$}\\
\multicolumn{5}{l}{$^c$ is 0.1--100 keV unabsorbed bolometric flux in units of $10^{-12}$ erg cm$^{-2}$ s$^{-1}$}\\
\end{tabular*}
\end{table*}

\begin{figure*}
   \includegraphics[width=\columnwidth]{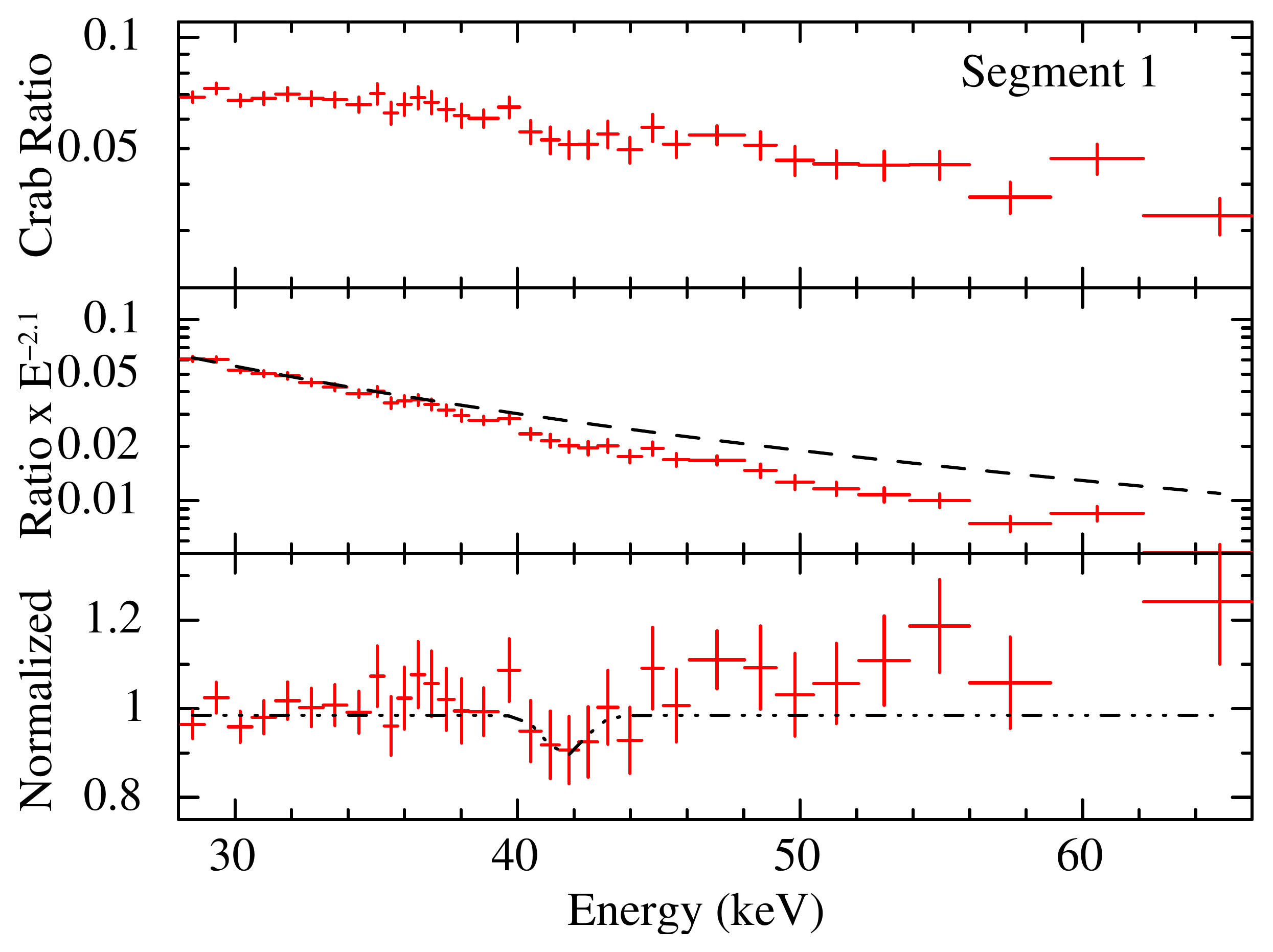}
   \includegraphics[width=\columnwidth]{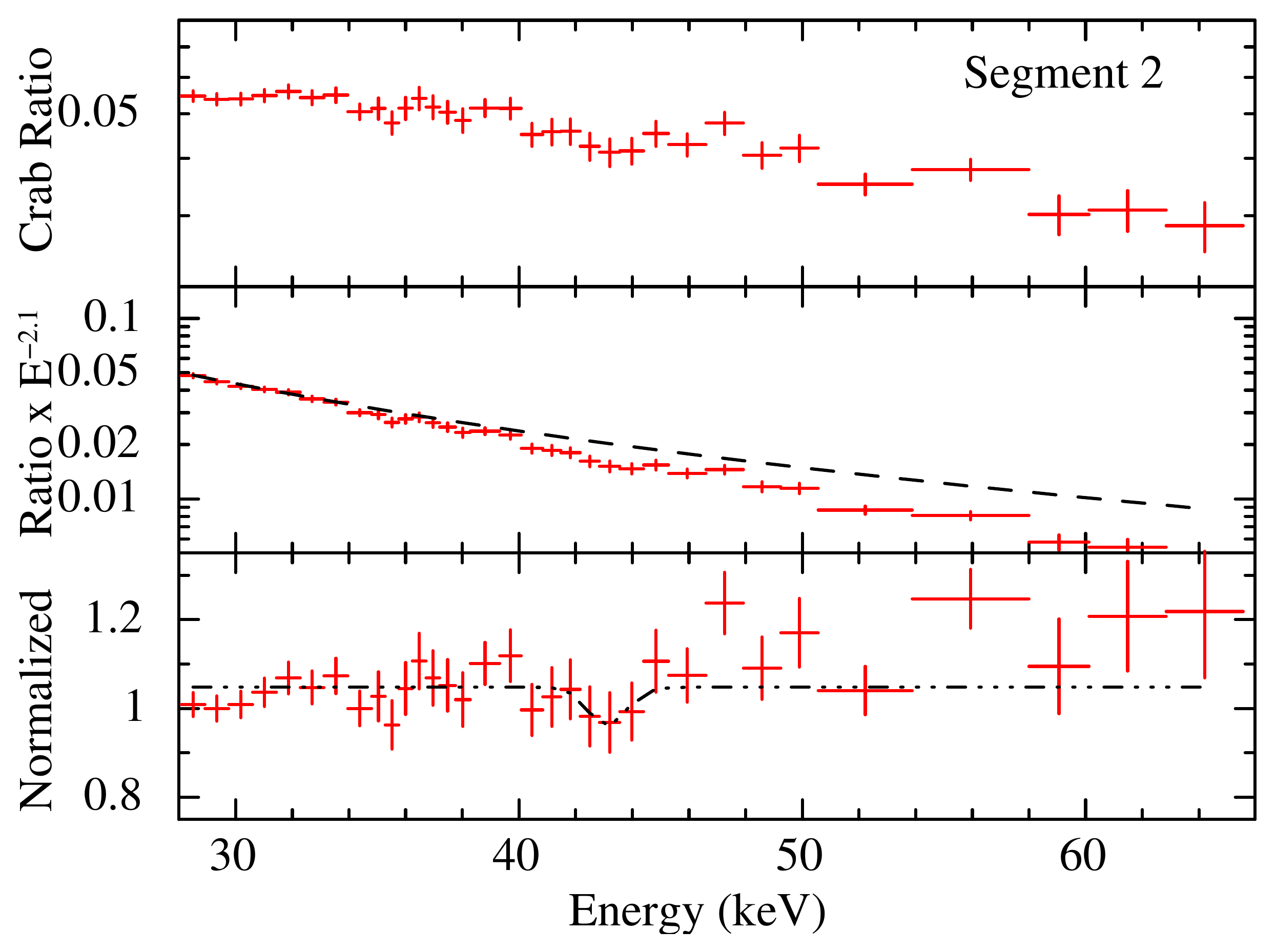}
 \centering
   \includegraphics[width=\columnwidth]{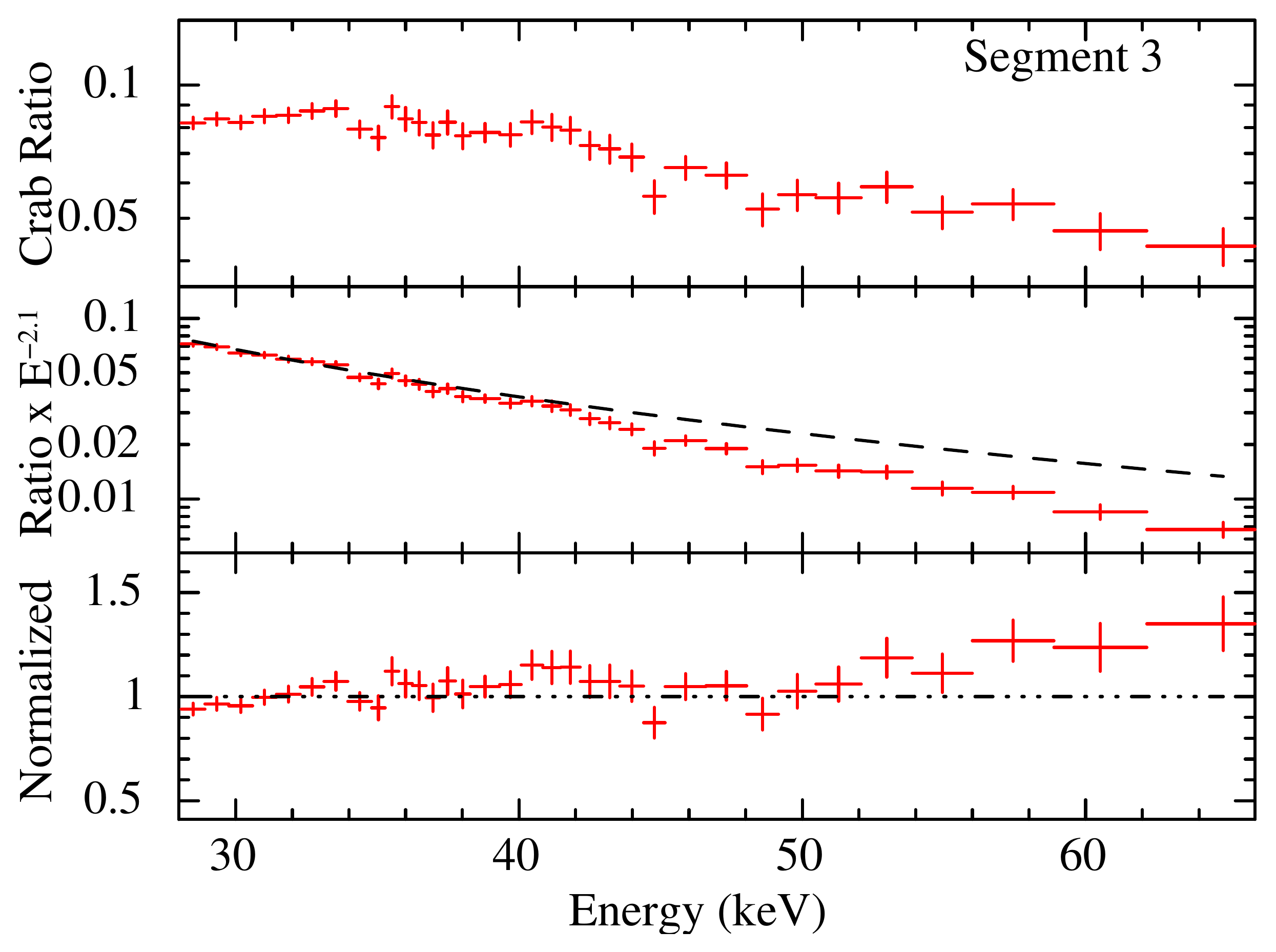}
\caption{Normalized Crab Ratio for OAO 1657-415. Top: Ratio between the OAO 1657-415 and Crab pulsar count rate spectra. Middle: Crab ratio multiplied by functional form of Crab spectrum (E$^{-2.1}$). Bottom: Normalized ratio after dividing the previous ratio by the continuum model of OAO 1657-415 derived from broad-band spectral fitting.}
 \label{fig:crab}
\end{figure*}

In order to characterize the CRSF better, we used a more reliable method of Normalized Crab ratio, extensively used in the past for several X-ray pulsars \citep{Mihara1995,Dal1998,Santangelo1998,Dal2000,Orlandini1998,Orlandini1999}. The spectrum of Crab pulsar is featureless above 10 keV. Therefore, the ratio between the source and Crab count rate spectra can be used to enhance the presence of features in the source spectrum. This method provides a model-independent analysis and minimizes effects due to instrument uncertainties. We used the \emph{NuSTAR} observation of Crab from August 2019 (Obs\_ID 10502001011). The top panels of Figure \ref{fig:crab} show the Crab ratio for each segment. For the first and second segments, a deviation from the monotonically decreasing trend around 40 keV was observed. We then multiplied the count rate ratio by functional form of Crab spectrum (middle panel Figure \ref{fig:crab}), a power law with an index of 2.1 \citep{Dal2000}, and divided it by the OAO 1657-415 continuum form described by our best-fitting model without the CRSF (Table \ref{tab:model}) to get the normalized ratio. The last panels of Figure \ref{fig:crab} show a narrow absorption feature centred at $\sim 40$ keV for segments 1 and 2, consistent with the best-fitting model. We did not find any evidence of CRSF in the spectrum from the third segment.  
 
We used \texttt{cflux} convolution model to calculate the unabsorbed flux in the energy range 0.1--100 keV for the complete model as well as its individual components. The emission line flux was comparable during the high flux intervals, segments 1 and 3 whereas it was higher during the relatively low flux segment 2. The total estimated flux decreased slightly from the first segment to the second and was highest during the last segment. 

\section{Discussion}

We have performed a comprehensive timing and spectral analysis of accretion powered HMXB pulsar OAO 1657-415 with \emph{NuSTAR} archival data from June 2019 and \emph{Fermi}/GBM data from October 2011 to July 2021. The \emph{NuSTAR} observation covered the orbital phase range of 0.57–0.74, where phase zero corresponds to the mid-eclipse \citep{Falanga2015}. OAO 1657-415 shows variation in the X-ray intensity by a factor of 3. We have resolved the entire observation into three segments, based on the HR, and studied the evolution of the timing and spectral properties during these intervals. We report the presence of Compton shoulder and cyclotron absorption like feature in the phase-averaged spectra of the source along with a Fe K$_{\alpha}$ emission line. We have also studied the long-term period evolution of the OAO 1657-415 covering 43 yr of spin period history. 

The X-ray intensity was relatively higher during the first and last segments as compared to the second segment. We detected X-ray pulsations in all three segments up to 70 keV. The estimated pulse periods indicate that OAO 1657-415 is undergoing a spin-down phase (Table \ref{tab:period}). We calculated a spin-down rate with $\dot{P} = 9(1) \times 10^{-8}$ s s$^{-1}$ for the entire observation.

We have studied the pulse profile of the source and found that the profile evolved with time as well as energy. The dip-like structure present at lower energies disappears at higher energies. Similar energy-dependent dip-like features at certain pulse phases have been reported in several HMXBs \citep{Tsy2007,Naik2011,Maitra2012,Naik2013}. The presence of inhomogenous stellar winds from the companion can trigger absorption of X-ray photons at lower energies resulting in dip-like structures in the pulse profiles \citep{Jaiswal2013}. 

We have studied the pulse period history over a span of 43 years. This span shows two characteristic features. The first one during MJD 43755 to MJD 56000 corresponds to long-term spin-up with short-term fluctuations occurring on time-scale of 1744 days. Such long-term period trends with periodic evolution of spin period have been observed in other X-ray pulsars also \citep{Gonzalez2012,Molkov2017,Vasilopoulos2019,Chandra2021}. Her X-1 \citep{Staubert2006} and GX 1+4 \citep{Gonzalez2012} also show similar period evolution with time. The short-term fluctuations in the period trend are interpreted to be due to the variation in the accretion rate on short time-scales causing reversals in the torque \citep{Ghosh1979,Nagase1989}. The possibility of such changes in the stellar wind activity of companion star has been proposed for the periodic variations of 9.2 yr in Cen X-3 \citep{Tsunemi1989}. Numerical simulations for wind-fed systems suggest that the specific angular momentum of the accreting mass can also lead to such short-term alternating trends of spin period \citep{Mitsuda1991,Foglizzo2005}.

A torque reversal from spin-up to spin-down occurred around MJD 57300. The rate of change of period is about an order of magnitude lower as compared to the spin-up phase with an average $\dot{P} \approx 8 \times 10^{-10}$ s s$^{-1}$ (for a linear fitting). A similar permanent reversal has been observed in Vela X-1 and GX 1+4 \citep{Tsunemi1989,Gonzalez2012}. For Vela X-1, \citet{Hayakawa1982} showed that the long-term changes in the stellar wind because of the variation in the activity of companion star can lead to such torque reversal. The absence of any correlation between X-ray flux and spin-down rate in GX 1+4 ruled out the possibility of a prograde or retrograde accretion disc for the observed torque reversal \citet{Gonzalez2012}. \citet{Jenke2012} studied the flux and torque correlation for OAO 1657-415 and suggested that during spin-down episodes the source exhibited no correlation with flux. While a retrograde disc could possibly explain the overall spinning down of OAO 1657-415 after MJD 57300, the presence of overlying spin-up episodes at short time-scales can not be justified. Moreover, the formation and stability of such a retrograde disc around a pulsar showing long-term spin-up for 37 yr can not be explained. Thus, similar to Vela X-1, the variation in the stellar wind from the companion star can explain the permanent torque reversal in OAO 1657-415.

We have performed the broad-band spectral analysis of OAO 1657-415. The time-resolved spectra of the source are described with a partially covered cutoff power law. The best-fitting spectral parameters are consistent with the highly obscured nature of OAO 1657-415 \citep{Orlandini1999,Jaiswal2014,Audley2006}. The column density varies across the segments, possibly indicating changes in the absorption intrinsic to the source, because of the presence of clumpy obscuring matter at late orbital phases \citep{Pradhan2014}. The covering fraction from the partial covering component is fairly consistent across three segments. We detect a modest softening of the spectrum as $\Gamma$ increases from $0.49_{-0.04}^{+0.08}$ during the relatively variable first segment to $0.80_{-0.03}^{+0.02}$ for the second segment. Spectrum becomes harder during the last segment.    

Presence of multiple emission lines has been detected in the spectrum of OAO 1657-415 with Fe K$_{\alpha}$ line as the most abundant \citep{Kamata1990,Orlandini1999,Jaiswal2014,Pradhan2014,Pradhan2019}. We detect Fe K$_{\alpha}$ emission line in the spectra from all three segments. The line flux varies across the three segments with the highest during the second segment. The equivalent width of  the emission line during second segment is significantly higher compared to the other two segments. The high column density, low flux, and large equivalent width during the second segment are indicative of the presence of dense matter cloud surrounding the source \citep{Pradhan2014,Pradhan2019}. The variation of the equivalent width with absorption column density is typical for X-ray sources partially obscured with absorbing matter distributed inhomogeneously \citep{Jaiswal2015}. The increase in the net flux during the last segment possibly results from the increased accretion rate as more matter is captured during passage through the dense matter clump \citep{Pradhan2014}. We do not find any emission lines from highly ionized species of Fe in the spectra from three segments implying a low degree of ionization during the observation \citep[$\xi \leq 10^2$; ][]{Kallman1982}.

The reprocessing of Fe K$_{\alpha}$ emission line photons in an optically thick dense cloud can result in a Compton shoulder between 6.24 and 6.40 keV \citep{Watanabe2003}. We detect the presence of an emission like feature at $6.36_{-0.02}^{+0.01}, 6.34 \pm 0.01$, and $6.32_{-0.01}^{+0.11}$ keV for three segments, respectively, along with Fe K$_{\alpha}$ emission line. We identify this feature as a Compton shoulder following the detection of a similar feature around 6.3 keV in the high-resolution \emph{Chandra} spectrum of OAO 1657-415 by \citet{Pradhan2019}. It is worth noticing here that previous works have tied the presence of dense clump of matter with non-detection of pulsations and presence of Compton shoulder in observations made at an orbital phase range of 0.5-0.6. The observation used in this work also covered the same orbital phase. We report confirmed presence of pulsations and Compton shoulder in our analysis. 

We have found a CRSF in the spectra from the first and second segments. We have modelled this feature by using \texttt{gabs} and \texttt{cyclabs} components. Both the models provide identically good fitting with consistent best-fitting parameters (Table \ref{tab:model}). The possible detection of a similar feature around 36 keV in the spectrum of OAO 1657-415 have been reported earlier also. But all the previous works are inconclusive either due to limited statistics of the data or due to poorly constrained parameters. We use the model-independent method of Normalized Crab ratio to ascertain the presence of this feature (Figure \ref{fig:crab}). The detection of such CRSF can provide a direct estimate of the magnetic field strength of the NS with the centroid energy of the feature related to the magnetic field strength as,

\begin{equation}
 E_{\rm CRSF} = 11.6 \ \rm keV \times (1+z)^{-1} \times \frac{B}{10^{12} \ \rm G}
\label{eq:tau}
\end{equation}

where $z$ is the gravitational redshift. We estimate a magnetic field strength of $3.59 \pm 0.06 \ (1 + z)^{-1} \times 10^{12}$ G and $3.29_{-0.22}^{+0.23} \ (1 + z)^{-1} \times 10^{12}$ G for the best-fitting CRSF line energies from the first and second segments, respectively (for results with \texttt{gabs}). The results are fairly consistent within error for the two segments.

\section*{Acknowledgements}

We have used the data obtained with the space mission \emph{NuSTAR}, a NASA funded project led by the California Institute of Technology. We have used the data archived by the High Energy Astrophysics Science Archive Research Center (HEASARC) online service maintained by the Goddard Space Flight Center. This work has made use of the \emph{NuSTAR} Data Analysis Software (NuSTARDAS) jointly developed by the ASI Space Science Data Center (SSDC, Italy) and the California Institute of Technology (Caltech, USA). We thank the referee for the valuable suggestions which have helped to improve this paper. AD acknowledges the support received from the IoE, University of Delhi as FRP grant. PS acknowledges the financial support from the Council of Scientific \& Industrial Research (CSIR) under the Junior Research Fellowship (JRF) scheme. 


\section*{Data Availability}
The \emph{NuSTAR} data underlying this article are available in the High Energy Astrophysics Science Archive Research Center (HEASARC) online service at \url{https://heasarc.gsfc.nasa.gov/cgi-bin/W3Browse/w3browse.pl}.
The \emph{CGRO}/BATSE spin period history for OAO 1657-415 is available at \url{https://gammaray.nsstc.nasa.gov/batse/pulsar/data/archive.html}.
The \emph{Fermi}/GBM period evolution history for OAO 1657-415 is available at \url{https://gammaray.nsstc.nasa.gov/gbm/science/pulsars/lightcurves/oao1657.html}.
 



\bibliographystyle{mnras}
\bibliography{OAO1657} 

\begin{thebibliography}{}
\makeatletter
\relax
\def\mn@urlcharsother{\let\do\@makeother \do\$\do\&\do\#\do\^\do\_\do\%\do\~}
\def\mn@doi{\begingroup\mn@urlcharsother \@ifnextchar [ {\mn@doi@}
  {\mn@doi@[]}}
\def\mn@doi@[#1]#2{\def\@tempa{#1}\ifx\@tempa\@empty \href
  {http://dx.doi.org/#2} {doi:#2}\else \href {http://dx.doi.org/#2} {#1}\fi
  \endgroup}
\def\mn@eprint#1#2{\mn@eprint@#1:#2::\@nil}
\def\mn@eprint@arXiv#1{\href {http://arxiv.org/abs/#1} {{\tt arXiv:#1}}}
\def\mn@eprint@dblp#1{\href {http://dblp.uni-trier.de/rec/bibtex/#1.xml}
  {dblp:#1}}
\def\mn@eprint@#1:#2:#3:#4\@nil{\def\@tempa {#1}\def\@tempb {#2}\def\@tempc
  {#3}\ifx \@tempc \@empty \let \@tempc \@tempb \let \@tempb \@tempa \fi \ifx
  \@tempb \@empty \def\@tempb {arXiv}\fi \@ifundefined
  {mn@eprint@\@tempb}{\@tempb:\@tempc}{\expandafter \expandafter \csname
  mn@eprint@\@tempb\endcsname \expandafter{\@tempc}}}

\bibitem[\protect\citeauthoryear{{Arnaud}}{{Arnaud}}{1996}]{Arnaud1996}
{Arnaud} K.~A.,  1996, in {Jacoby} G.~H.,  {Barnes} J.,  eds,  Astronomical
  Society of the Pacific Conference Series Vol. 101, Astronomical Data Analysis
  Software and Systems V. p.~17

\bibitem[\protect\citeauthoryear{{Audley}, {Nagase}, {Mitsuda}, {Angelini}  \&
  {Kelley}}{{Audley} et~al.}{2006}]{Audley2006}
{Audley} M.~D.,  {Nagase} F.,  {Mitsuda} K.,  {Angelini} L.,   {Kelley} R.~L.,
  2006, \mn@doi [\mnras] {10.1111/j.1365-2966.2006.10003.x}, \href
  {https://ui.adsabs.harvard.edu/abs/2006MNRAS.367.1147A} {367, 1147}

\bibitem[\protect\citeauthoryear{{Barnstedt} et~al.,}{{Barnstedt}
  et~al.}{2008}]{Barnstedt2008}
{Barnstedt} J.,  et~al., 2008, \mn@doi [\aap] {10.1051/0004-6361:20078707},
  \href {https://ui.adsabs.harvard.edu/abs/2008A&A...486..293B} {486, 293}

\bibitem[\protect\citeauthoryear{{Baykal}}{{Baykal}}{1997}]{Baykal1997}
{Baykal} A.,  1997, \aap, \href
  {https://ui.adsabs.harvard.edu/abs/1997A&A...319..515B} {319, 515}

\bibitem[\protect\citeauthoryear{{Baykal}}{{Baykal}}{2000}]{Baykal2000}
{Baykal} A.,  2000, \mn@doi [\mnras] {10.1046/j.1365-8711.2000.03249.x}, \href
  {https://ui.adsabs.harvard.edu/abs/2000MNRAS.313..637B} {313, 637}

\bibitem[\protect\citeauthoryear{{Becker} \& {Wolff}}{{Becker} \&
  {Wolff}}{2005}]{Becker2005}
{Becker} P.~A.,  {Wolff} M.~T.,  2005, \mn@doi [\apj] {10.1086/431720}, \href
  {https://ui.adsabs.harvard.edu/abs/2005ApJ...630..465B} {630, 465}

\bibitem[\protect\citeauthoryear{{Bildsten} et~al.,}{{Bildsten}
  et~al.}{1997}]{Bildsten1997}
{Bildsten} L.,  et~al., 1997, \mn@doi [\apjs] {10.1086/313060}, \href
  {https://ui.adsabs.harvard.edu/abs/1997ApJS..113..367B} {113, 367}

\bibitem[\protect\citeauthoryear{{Chakrabarty} et~al.,}{{Chakrabarty}
  et~al.}{1993}]{Chakrabarty1993}
{Chakrabarty} D.,  et~al., 1993, \mn@doi [\apjl] {10.1086/186715}, \href
  {https://ui.adsabs.harvard.edu/abs/1993ApJ...403L..33C} {403, L33}

\bibitem[\protect\citeauthoryear{{Chakrabarty}, {Wang}, {Juett}, {Lee}  \&
  {Roche}}{{Chakrabarty} et~al.}{2002}]{Chakrabarty2002}
{Chakrabarty} D.,  {Wang} Z.,  {Juett} A.~M.,  {Lee} J.~C.,   {Roche} P.,
  2002, \mn@doi [\apj] {10.1086/340746}, \href
  {https://ui.adsabs.harvard.edu/abs/2002ApJ...573..789C} {573, 789}

\bibitem[\protect\citeauthoryear{{Coburn}, {Heindl}, {Rothschild}, {Gruber},
  {Kreykenbohm}, {Wilms}, {Kretschmar}  \& {Staubert}}{{Coburn}
  et~al.}{2002}]{Coburn2002}
{Coburn} W.,  {Heindl} W.~A.,  {Rothschild} R.~E.,  {Gruber} D.~E.,
  {Kreykenbohm} I.,  {Wilms} J.,  {Kretschmar} P.,   {Staubert} R.,  2002,
  \mn@doi [\apj] {10.1086/343033}, \href
  {https://ui.adsabs.harvard.edu/abs/2002ApJ...580..394C} {580, 394}

\bibitem[\protect\citeauthoryear{{Corbet}}{{Corbet}}{1986}]{Corbet1986}
{Corbet} R.~H.~D.,  1986, \mn@doi [\mnras] {10.1093/mnras/220.4.1047}, \href
  {https://ui.adsabs.harvard.edu/abs/1986MNRAS.220.1047C} {220, 1047}

\bibitem[\protect\citeauthoryear{{Cutri} et~al.,}{{Cutri}
  et~al.}{2003}]{Cat2003}
{Cutri} R.~M.,  et~al., 2003, VizieR Online Data Catalog, \href
  {https://ui.adsabs.harvard.edu/abs/2003yCat.2246....0C} {p. II/246}

\bibitem[\protect\citeauthoryear{{Deeter}, {Boynton}, {Lamb}  \&
  {Zylstra}}{{Deeter} et~al.}{1989}]{Deeter1989}
{Deeter} J.~E.,  {Boynton} P.~E.,  {Lamb} F.~K.,   {Zylstra} G.,  1989, \mn@doi
  [\apj] {10.1086/167017}, \href
  {https://ui.adsabs.harvard.edu/abs/1989ApJ...336..376D} {336, 376}

\bibitem[\protect\citeauthoryear{{Deo Chandra}, {Roy}, {Agrawal}  \&
  {Choudhury}}{{Deo Chandra} et~al.}{2021}]{Chandra2021}
{Deo Chandra} A.,  {Roy} J.,  {Agrawal} P.~C.,   {Choudhury} M.,  2021, arXiv
  e-prints, \href {https://ui.adsabs.harvard.edu/abs/2021arXiv210807097D} {p.
  arXiv:2108.07097}

\bibitem[\protect\citeauthoryear{{Falanga}, {Bozzo}, {Lutovinov},
  {Bonnet-Bidaud}, {Fetisova}  \& {Puls}}{{Falanga} et~al.}{2015}]{Falanga2015}
{Falanga} M.,  {Bozzo} E.,  {Lutovinov} A.,  {Bonnet-Bidaud} J.~M.,  {Fetisova}
  Y.,   {Puls} J.,  2015, \mn@doi [\aap] {10.1051/0004-6361/201425191}, \href
  {https://ui.adsabs.harvard.edu/abs/2015A&A...577A.130F} {577, A130}

\bibitem[\protect\citeauthoryear{{Foglizzo}, {Galletti}  \&
  {Ruffert}}{{Foglizzo} et~al.}{2005}]{Foglizzo2005}
{Foglizzo} T.,  {Galletti} P.,   {Ruffert} M.,  2005, \mn@doi [\aap]
  {10.1051/0004-6361:20042201}, \href
  {https://ui.adsabs.harvard.edu/abs/2005A&A...435..397F} {435, 397}

\bibitem[\protect\citeauthoryear{{Ghosh} \& {Lamb}}{{Ghosh} \&
  {Lamb}}{1979}]{Ghosh1979}
{Ghosh} P.,  {Lamb} F.~K.,  1979, \mn@doi [\apj] {10.1086/157498}, \href
  {https://ui.adsabs.harvard.edu/abs/1979ApJ...234..296G} {234, 296}

\bibitem[\protect\citeauthoryear{{Gonz{\'a}lez-Gal{\'a}n}, {Kuulkers},
  {Kretschmar}, {Larsson}, {Postnov}, {Kochetkova}  \&
  {Finger}}{{Gonz{\'a}lez-Gal{\'a}n} et~al.}{2012}]{Gonzalez2012}
{Gonz{\'a}lez-Gal{\'a}n} A.,  {Kuulkers} E.,  {Kretschmar} P.,  {Larsson} S.,
  {Postnov} K.,  {Kochetkova} A.,   {Finger} M.~H.,  2012, \mn@doi [\aap]
  {10.1051/0004-6361/201117893}, \href
  {https://ui.adsabs.harvard.edu/abs/2012A&A...537A..66G} {537, A66}

\bibitem[\protect\citeauthoryear{{Harrison} et~al.,}{{Harrison}
  et~al.}{2013}]{Harrison2013}
{Harrison} F.~A.,  et~al., 2013, \mn@doi [\apj] {10.1088/0004-637X/770/2/103},
  \href {https://ui.adsabs.harvard.edu/abs/2013ApJ...770..103H} {770, 103}

\bibitem[\protect\citeauthoryear{{Hayakawa}}{{Hayakawa}}{1982}]{Hayakawa1982}
{Hayakawa} S.,  1982, in {Brinkmann} W.,  {Truemper} J.,  eds, Accreting
  Neutron Stars. pp 14--28

\bibitem[\protect\citeauthoryear{{Jaisawal} \& {Naik}}{{Jaisawal} \&
  {Naik}}{2014}]{Jaiswal2014}
{Jaisawal} G.~K.,  {Naik} S.,  2014, Bulletin of the Astronomical Society of
  India, \href {https://ui.adsabs.harvard.edu/abs/2014BASI...42..147J} {42,
  147}

\bibitem[\protect\citeauthoryear{{Jaisawal} \& {Naik}}{{Jaisawal} \&
  {Naik}}{2015}]{Jaiswal2015}
{Jaisawal} G.~K.,  {Naik} S.,  2015, \mn@doi [\mnras] {10.1093/mnras/stv029},
  \href {https://ui.adsabs.harvard.edu/abs/2015MNRAS.448..620J} {448, 620}

\bibitem[\protect\citeauthoryear{{Jaisawal}, {Naik}  \& {Paul}}{{Jaisawal}
  et~al.}{2013}]{Jaiswal2013}
{Jaisawal} G.~K.,  {Naik} S.,   {Paul} B.,  2013, \mn@doi [\apj]
  {10.1088/0004-637X/779/1/54}, \href
  {https://ui.adsabs.harvard.edu/abs/2013ApJ...779...54J} {779, 54}

\bibitem[\protect\citeauthoryear{{Jaisawal}, {Naik}, {Epili}, {Chhotaray},
  {Jana}  \& {Agrawal}}{{Jaisawal} et~al.}{2021}]{Jaiswal2021}
{Jaisawal} G.~K.,  {Naik} S.,  {Epili} P.~R.,  {Chhotaray} B.,  {Jana} A.,
  {Agrawal} P.~C.,  2021, \mn@doi [Journal of Astrophysics and Astronomy]
  {10.1007/s12036-021-09701-x}, \href
  {https://ui.adsabs.harvard.edu/abs/2021JApA...42...72J} {42, 72}

\bibitem[\protect\citeauthoryear{{Jenke}, {Finger}, {Wilson-Hodge}  \&
  {Camero-Arranz}}{{Jenke} et~al.}{2012}]{Jenke2012}
{Jenke} P.~A.,  {Finger} M.~H.,  {Wilson-Hodge} C.~A.,   {Camero-Arranz} A.,
  2012, \mn@doi [\apj] {10.1088/0004-637X/759/2/124}, \href
  {https://ui.adsabs.harvard.edu/abs/2012ApJ...759..124J} {759, 124}

\bibitem[\protect\citeauthoryear{{Kallman} \& {McCray}}{{Kallman} \&
  {McCray}}{1982}]{Kallman1982}
{Kallman} T.~R.,  {McCray} R.,  1982, \mn@doi [\apjs] {10.1086/190828}, \href
  {https://ui.adsabs.harvard.edu/abs/1982ApJS...50..263K} {50, 263}

\bibitem[\protect\citeauthoryear{{Kamata}, {Koyama}, {Tawara}, {Makishima},
  {Ohashi}, {Kawai}  \& {Hatsukade}}{{Kamata} et~al.}{1990}]{Kamata1990}
{Kamata} Y.,  {Koyama} K.,  {Tawara} Y.,  {Makishima} K.,  {Ohashi} T.,
  {Kawai} N.,   {Hatsukade} I.,  1990, \pasj, \href
  {https://ui.adsabs.harvard.edu/abs/1990PASJ...42..785K} {42, 785}

\bibitem[\protect\citeauthoryear{{Maitra}, {Paul}  \& {Naik}}{{Maitra}
  et~al.}{2012}]{Maitra2012}
{Maitra} C.,  {Paul} B.,   {Naik} S.,  2012, \mn@doi [\mnras]
  {10.1111/j.1365-2966.2011.20196.x}, \href
  {https://ui.adsabs.harvard.edu/abs/2012MNRAS.420.2307M} {420, 2307}

\bibitem[\protect\citeauthoryear{{Mason}, {Clark}, {Norton}, {Negueruela}  \&
  {Roche}}{{Mason} et~al.}{2009}]{Mason2009}
{Mason} A.~B.,  {Clark} J.~S.,  {Norton} A.~J.,  {Negueruela} I.,   {Roche} P.,
   2009, \mn@doi [\aap] {10.1051/0004-6361/200912480}, \href
  {https://ui.adsabs.harvard.edu/abs/2009A&A...505..281M} {505, 281}

\bibitem[\protect\citeauthoryear{{Matsuda}, {Sekino}, {Sawada}, {Shima},
  {Livio}, {Anzer}  \& {Boerner}}{{Matsuda} et~al.}{1991}]{Mitsuda1991}
{Matsuda} T.,  {Sekino} N.,  {Sawada} K.,  {Shima} E.,  {Livio} M.,  {Anzer}
  U.,   {Boerner} G.,  1991, \aap, \href
  {https://ui.adsabs.harvard.edu/abs/1991A&A...248..301M} {248, 301}

\bibitem[\protect\citeauthoryear{{Meegan} et~al.,}{{Meegan}
  et~al.}{2009}]{Meegan2009}
{Meegan} C.,  et~al., 2009, \mn@doi [\apj] {10.1088/0004-637X/702/1/791}, \href
  {https://ui.adsabs.harvard.edu/abs/2009ApJ...702..791M} {702, 791}

\bibitem[\protect\citeauthoryear{{Mereghetti} et~al.,}{{Mereghetti}
  et~al.}{1991}]{Mereghetti1991}
{Mereghetti} S.,  et~al., 1991, \mn@doi [\apjl] {10.1086/185901}, \href
  {https://ui.adsabs.harvard.edu/abs/1991ApJ...366L..23M} {366, L23}

\bibitem[\protect\citeauthoryear{{Mihara}}{{Mihara}}{1995}]{Mihara1995}
{Mihara} T.,  1995, PhD thesis, -

\bibitem[\protect\citeauthoryear{{Molkov}, {Lutovinov}, {Falanga}, {Tsygankov}
  \& {Bozzo}}{{Molkov} et~al.}{2017}]{Molkov2017}
{Molkov} S.,  {Lutovinov} A.,  {Falanga} M.,  {Tsygankov} S.,   {Bozzo} E.,
  2017, \mn@doi [\mnras] {10.1093/mnras/stw2429}, \href
  {https://ui.adsabs.harvard.edu/abs/2017MNRAS.464.2039M} {464, 2039}

\bibitem[\protect\citeauthoryear{{Nagase}}{{Nagase}}{1989}]{Nagase1989}
{Nagase} F.,  1989, \pasj, \href
  {https://ui.adsabs.harvard.edu/abs/1989PASJ...41....1N} {41, 1}

\bibitem[\protect\citeauthoryear{{Nagase} et~al.,}{{Nagase}
  et~al.}{1984}]{Nagase1984}
{Nagase} F.,  et~al., 1984, \pasj, \href
  {https://ui.adsabs.harvard.edu/abs/1984PASJ...36..667N} {36, 667}

\bibitem[\protect\citeauthoryear{{Naik}, {Paul}, {Kachhara}  \&
  {Vadawale}}{{Naik} et~al.}{2011}]{Naik2011}
{Naik} S.,  {Paul} B.,  {Kachhara} C.,   {Vadawale} S.~V.,  2011, \mn@doi
  [\mnras] {10.1111/j.1365-2966.2010.18128.x}, \href
  {https://ui.adsabs.harvard.edu/abs/2011MNRAS.413..241N} {413, 241}

\bibitem[\protect\citeauthoryear{{Naik}, {Maitra}, {Jaisawal}  \&
  {Paul}}{{Naik} et~al.}{2013}]{Naik2013}
{Naik} S.,  {Maitra} C.,  {Jaisawal} G.~K.,   {Paul} B.,  2013, \mn@doi [\apj]
  {10.1088/0004-637X/764/2/158}, \href
  {https://ui.adsabs.harvard.edu/abs/2013ApJ...764..158N} {764, 158}

\bibitem[\protect\citeauthoryear{{Orlandini} et~al.,}{{Orlandini}
  et~al.}{1998}]{Orlandini1998}
{Orlandini} M.,  et~al., 1998, \aap, \href
  {https://ui.adsabs.harvard.edu/abs/1998A&A...332..121O} {332, 121}

\bibitem[\protect\citeauthoryear{{Orlandini}, {dal Fiume}, {del Sordo},
  {Frontera}, {Parmar}, {Santangelo}  \& {Segreto}}{{Orlandini}
  et~al.}{1999}]{Orlandini1999}
{Orlandini} M.,  {dal Fiume} D.,  {del Sordo} S.,  {Frontera} F.,  {Parmar}
  A.~N.,  {Santangelo} A.,   {Segreto} A.,  1999, \aap, \href
  {https://ui.adsabs.harvard.edu/abs/1999A&A...349L...9O} {349, L9}

\bibitem[\protect\citeauthoryear{{Parmar} et~al.,}{{Parmar}
  et~al.}{1980}]{Parmar1980}
{Parmar} A.~N.,  et~al., 1980, \mn@doi [\mnras] {10.1093/mnras/193.1.49P},
  \href {https://ui.adsabs.harvard.edu/abs/1980MNRAS.193P..49P} {193, 49P}

\bibitem[\protect\citeauthoryear{{Polidan}, {Pollard}, {Sanford}  \&
  {Locke}}{{Polidan} et~al.}{1978}]{Polidan1978}
{Polidan} R.~S.,  {Pollard} G.~S.~G.,  {Sanford} P.~W.,   {Locke} M.~C.,  1978,
  \mn@doi [\nat] {10.1038/275296a0}, \href
  {https://ui.adsabs.harvard.edu/abs/1978Natur.275..296P} {275, 296}

\bibitem[\protect\citeauthoryear{{Pradhan}, {Maitra}, {Paul}, {Islam}  \&
  {Paul}}{{Pradhan} et~al.}{2014}]{Pradhan2014}
{Pradhan} P.,  {Maitra} C.,  {Paul} B.,  {Islam} N.,   {Paul} B.~C.,  2014,
  \mn@doi [\mnras] {10.1093/mnras/stu1034}, \href
  {https://ui.adsabs.harvard.edu/abs/2014MNRAS.442.2691P} {442, 2691}

\bibitem[\protect\citeauthoryear{{Pradhan}, {Raman}  \& {Paul}}{{Pradhan}
  et~al.}{2019}]{Pradhan2019}
{Pradhan} P.,  {Raman} G.,   {Paul} B.,  2019, \mn@doi [\mnras]
  {10.1093/mnras/sty3441}, \href
  {https://ui.adsabs.harvard.edu/abs/2019MNRAS.483.5687P} {483, 5687}

\bibitem[\protect\citeauthoryear{{Reig}}{{Reig}}{2011}]{Reig2011}
{Reig} P.,  2011, \mn@doi [\apss] {10.1007/s10509-010-0575-8}, \href
  {https://ui.adsabs.harvard.edu/abs/2011Ap&SS.332....1R} {332, 1}

\bibitem[\protect\citeauthoryear{{Santangelo}, {del Sordo}, {Segreto}, {dal
  Fiume}, {Orlandini}  \& {Piraino}}{{Santangelo}
  et~al.}{1998}]{Santangelo1998}
{Santangelo} A.,  {del Sordo} S.,  {Segreto} A.,  {dal Fiume} D.,  {Orlandini}
  M.,   {Piraino} S.,  1998, \aap, \href
  {https://ui.adsabs.harvard.edu/abs/1998A&A...340L..55S} {340, L55}

\bibitem[\protect\citeauthoryear{{Staubert}, {Schandl}, {Klochkov}, {Wilms},
  {Postnov}  \& {Shakura}}{{Staubert} et~al.}{2006}]{Staubert2006}
{Staubert} R.,  {Schandl} S.,  {Klochkov} D.,  {Wilms} J.,  {Postnov} K.,
  {Shakura} N.,  2006, in {D'Amico} F.,  {Braga} J.,   {Rothschild} R.~E.,
  eds,  American Institute of Physics Conference Series Vol. 840, The Transient
  Milky Way: A Perspective for MIRAX. pp 65--70 (\mn@eprint {arXiv}
  {astro-ph/0702528}), \mn@doi{10.1063/1.2216605}

\bibitem[\protect\citeauthoryear{{Sunyaev}, {Gilfanov}, {Goldurm}  \&
  {Schmitz-Fraysse}}{{Sunyaev} et~al.}{1991}]{Sunyaev1991}
{Sunyaev} R.,  {Gilfanov} M.,  {Goldurm} A.,   {Schmitz-Fraysse} M.~C.,  1991,
  \iaucirc, \href {https://ui.adsabs.harvard.edu/abs/1991IAUC.5342....2S}
  {5342, 2}

\bibitem[\protect\citeauthoryear{{Tsunemi}}{{Tsunemi}}{1989}]{Tsunemi1989}
{Tsunemi} H.,  1989, \pasj, \href
  {https://ui.adsabs.harvard.edu/abs/1989PASJ...41..453T} {41, 453}

\bibitem[\protect\citeauthoryear{{Tsygankov}, {Lutovinov}, {Churazov}  \&
  {Sunyaev}}{{Tsygankov} et~al.}{2007}]{Tsy2007}
{Tsygankov} S.~S.,  {Lutovinov} A.~A.,  {Churazov} E.~M.,   {Sunyaev} R.~A.,
  2007, \mn@doi [Astronomy Letters] {10.1134/S1063773707060023}, \href
  {https://ui.adsabs.harvard.edu/abs/2007AstL...33..368T} {33, 368}

\bibitem[\protect\citeauthoryear{{Vasilopoulos}, {Petropoulou}, {Koliopanos},
  {Ray}, {Bailyn}, {Haberl}  \& {Gendreau}}{{Vasilopoulos}
  et~al.}{2019}]{Vasilopoulos2019}
{Vasilopoulos} G.,  {Petropoulou} M.,  {Koliopanos} F.,  {Ray} P.~S.,  {Bailyn}
  C.~B.,  {Haberl} F.,   {Gendreau} K.,  2019, \mn@doi [\mnras]
  {10.1093/mnras/stz2045}, \href
  {https://ui.adsabs.harvard.edu/abs/2019MNRAS.488.5225V} {488, 5225}

\bibitem[\protect\citeauthoryear{{Verner}, {Ferland}, {Korista}  \&
  {Yakovlev}}{{Verner} et~al.}{1996}]{Verner1996}
{Verner} D.~A.,  {Ferland} G.~J.,  {Korista} K.~T.,   {Yakovlev} D.~G.,  1996,
  \mn@doi [\apj] {10.1086/177435}, \href
  {https://ui.adsabs.harvard.edu/abs/1996ApJ...465..487V} {465, 487}

\bibitem[\protect\citeauthoryear{{Walter}, {Lutovinov}, {Bozzo}  \&
  {Tsygankov}}{{Walter} et~al.}{2015}]{Walter2015}
{Walter} R.,  {Lutovinov} A.~A.,  {Bozzo} E.,   {Tsygankov} S.~S.,  2015,
  \mn@doi [\aapr] {10.1007/s00159-015-0082-6}, \href
  {https://ui.adsabs.harvard.edu/abs/2015A&ARv..23....2W} {23, 2}

\bibitem[\protect\citeauthoryear{{Watanabe} et~al.,}{{Watanabe}
  et~al.}{2003}]{Watanabe2003}
{Watanabe} S.,  et~al., 2003, \mn@doi [\apjl] {10.1086/379735}, \href
  {https://ui.adsabs.harvard.edu/abs/2003ApJ...597L..37W} {597, L37}

\bibitem[\protect\citeauthoryear{{White} \& {Pravdo}}{{White} \&
  {Pravdo}}{1979}]{White1979}
{White} N.~E.,  {Pravdo} S.~H.,  1979, \mn@doi [\apjl] {10.1086/183089}, \href
  {https://ui.adsabs.harvard.edu/abs/1979ApJ...233L.121W} {233, L121}

\bibitem[\protect\citeauthoryear{{White}, {Swank}  \& {Holt}}{{White}
  et~al.}{1983}]{White1983}
{White} N.~E.,  {Swank} J.~H.,   {Holt} S.~S.,  1983, \mn@doi [\apj]
  {10.1086/161162}, \href
  {https://ui.adsabs.harvard.edu/abs/1983ApJ...270..711W} {270, 711}

\bibitem[\protect\citeauthoryear{{Wilms}, {Allen}  \& {McCray}}{{Wilms}
  et~al.}{2000}]{Wilms2000}
{Wilms} J.,  {Allen} A.,   {McCray} R.,  2000, \mn@doi [\apj] {10.1086/317016},
  \href {https://ui.adsabs.harvard.edu/abs/2000ApJ...542..914W} {542, 914}

\bibitem[\protect\citeauthoryear{{dal Fiume} et~al.,}{{dal Fiume}
  et~al.}{1998}]{Dal1998}
{dal Fiume} D.,  et~al., 1998, \aap, \href
  {https://ui.adsabs.harvard.edu/abs/1998A&A...329L..41D} {329, L41}

\bibitem[\protect\citeauthoryear{{dal Fiume} et~al.,}{{dal Fiume}
  et~al.}{2000}]{Dal2000}
{dal Fiume} D.,  et~al., 2000, \mn@doi [Advances in Space Research]
  {10.1016/S0273-1177(99)00767-X}, \href
  {https://ui.adsabs.harvard.edu/abs/2000AdSpR..25..399D} {25, 399}

\makeatother
\end{thebibliography}








\bsp	
\label{lastpage}
\end{document}